\documentclass[twocolumn]{emulateapj}

\newcommand{\Lsun}{$L_{\odot}$}
\newcommand{\Msun}{$M_{\odot}$}
\newcommand{\Rsun}{$R_{\odot}$}
\newcommand{\Mdot}{$\dot{M}$}

\shorttitle{Characterizing the stellar photospheres and near-infrared excesses in accreting T Tauri systems}
\shortauthors{McClure et al.}

\begin{document}

\title{Characterizing the stellar photospheres and near-infrared excesses in accreting T Tauri systems}


\author{M. K. McClure\altaffilmark{1,2,3}, N. Calvet\altaffilmark{1},  C. Espaillat\altaffilmark{4}, L. Hartmann\altaffilmark{1}, J. Hern\'andez\altaffilmark{5}, L. Ingleby\altaffilmark{1},K. L. Luhman\altaffilmark{6}, P. D'Alessio\altaffilmark{7}, B. Sargent\altaffilmark{8}}

\altaffiltext{1}{Department of Astronomy, The University of Michigan, 500 Church St., 830 Dennison Bldg., Ann Arbor, MI 48109; melisma@umich.edu, ncalvet@umich.edu, lhartm@umich.edu, lingleby@umich.edu}
\altaffiltext{2}{NSF Graduate Research Fellow}
\altaffiltext{3}{Visiting Astronomer at the Infrared Telescope Facility, which is operated by the University of Hawaii under Cooperative Agreement no. NNX-08AE38A with the National Aeronautics and Space Administration, Science Mission Directorate, Planetary Astronomy Program.}
\altaffiltext{4}{Center for Astrophysics, 60 Garden Street, Cambridge, MA 02138, USA; Sagan Fellow; cespaillat@cfa.harvard.edu}
\altaffiltext{5}{Centro de Investigaciones de Astronom\'ia (CIDA), M\'erida 5101-A, Venezuela; hernandj@cida.ve}
\altaffiltext{6}{Department of Astronomy and Astrophysics and the Center for Exoplanets and Habitable Worlds, The Pennsylvania State University, University Park, PA 16802, USA; kluhman@astro.psu.edu}
\altaffiltext{7}{Centro de Radioastronom\'{i}a y Astrof\'{i}sica, Universidad Nacional Aut\'{o}noma de M\'{e}xico, 58089 Morelia, Michoac\'{a}n, M\'{e}xico; p.dalessio@astrosmo.unam.mx}
\altaffiltext{8}{Center for Imaging Science and Laboratory for Multiwavelength Astrophysics, Rochester Institute of Technology, 54 Lomb Memorial Drive, Rochester, NY 14623, USA; baspci@rit.edu}

\begin{abstract}

Using NASA IRTF SpeX data from 0.8 to 4.5 $\mu$m, we determine self-consistently the stellar properties and excess emission above the photosphere for a sample of classical T Tauri stars (CTTS) in the Taurus molecular cloud with varying degrees of accretion.   This process uses a combination of techniques from the recent literature as well as observations of weak-line T Tauri stars (WTTS) to account for the differences in surface gravity and chromospheric activity between the TTS and dwarfs, which are typically used as photospheric templates for CTTS.  Our improved veiling and extinction estimates for our targets allow us to extract flux-calibrated spectra of the excess in the near-infrared.  We find that we are able to produce an acceptable parametric fit to the near-infrared excesses using a combination of up to three blackbodies.  In half of our sample, two blackbodies at temperatures of 8000 K and 1600 K suffice.  These temperatures and the corresponding solid angles are consistent with emission from the accretion shock on the stellar surface and the inner dust sublimation rim of the disk, respectively.  In contrast, the other half requires three blackbodies at 8000, 1800, and 800 K, to describe the excess.  We interpret the combined two cooler blackbodies as the dust sublimation wall with either a contribution from the disk surface beyond the wall or curvature of the wall itself, neither of which should have single-temperature blackbody emission.  In these fits, we find no evidence of a contribution from optically thick gas inside the inner dust rim. 

%We also propose that the  is suggestive of the wall atmosphere optical depth and grain size effects modeled by \citet{dalessio+05} and \citet{ajay+07}, respectively.

%, and make rough estimates of the temperatures, solid angles, radii, and heights of their dust sublimation walls.  
%appropriate to the dust sublimation temperature without requiring a contribution from a gaseous inner disk

\end{abstract}


\section{Introduction}
\label{intro}

It is now accepted, from studies of UV and optical spectroscopy and near-infrared (NIR) interferometry, that the excess emission shortward of 5 $\micron$ in spectral energy distributions (SEDs) of classical T Tauri stars (CTTS) arises mainly in two physical components.  The first, responsible for the near-infrared (NIR) excess, is the sharp inner edge, or `wall', of the dust disk at which the disk temperature becomes high enough to sublimate the dust \citep{natta+01}.  Inside the dust-sublimation radius only gas remains, spiraling in until it is accreted onto the star through magnetospheric accretion columns.  The second emission component, responsible for the UV and optical excess, is the accretion shock formed when material free-falling along stellar magnetic field lines merges with the stellar photosphere through a shock at the surface, where most of the accretion luminosity is emitted \citep{cg98}.  

One can constrain the accretion rate and the structure of the inner disk wall from the shape, absolute flux, and line emission seen in the excess.  However, extracting the excess from the observed spectrum requires subtracting the underlying stellar photosphere and correcting for extinction along the line of sight.  This is typically accomplished by identifying the photospheric template that best matches the intrinsic stellar spectrum of the program star, assuming that the weaker absorption features in the program star are due to the excess continuum emission filling in, or `veiling', these features.   By comparing the difference in feature depth or equivalent width between the program star and template it is possible to determine the fraction of the observed flux contributed by the stellar photosphere \citep{bb90,heg95,muzerolle03,espaillat+10}. The slopes of the veiling-corrected spectrum over a range of wavelengths can be compared with that of the intrinsic template to derive the extinction, $A_V$.  Subtraction of the intrinsic template, offset from the observed spectrum by the appropriate degree of veiling, from the extinction-corrected observed spectrum yields the veiling spectrum \citep{gullbring+98a,sd+08,fischer+11}.  If the derived excess spectrum is absolutely flux calibrated, e.g. with simultaneous photometry, then one also can obtain sizes of its emitting regions.

The NIR excess is of particular interest as it measures directly the inner disk emission, thus providing insight into the state of the inner disk and its effect on the outer disk.  For example, the maximum grain size and composition in the wall dust population affect its shape and hence the emitting area \citep{isellanatta05,ajay+07}.  There is also evidence for an optically thick gas component inside the dust sublimation radius \citep[][and references therein]{ajay+08, eisner+07, fischer+11}. The geometry of the wall relative to the disk behind it determines how much stellar emission is incident on the outer disk; if the wall `shadows' the disk it would prevent the middle few AU from being heated effectively, producing less flaring \citep{natta+01,dullemond+01, meeus+01,dd04}.  

In practice it can be difficult to successfully apply the standard approach for excess extraction.  Careful selection is required to avoid lines that show differential veiling due to chromospheric emission \citep{fb87,bb93}.  It is also essential to select the correct template star; the best choice is a weak-line T Tauri stars (WTTS), which does not accrete and has no NIR excess.  These stars are quantitatively distinguished from the CTTS by the strength of their $H\alpha$ emission lines \citep{wg01}. Of possible template stars, WTTS have the closest physical properties to the CTTS, including active chromospheres, comparable metallicities, and similar surface gravities.  The latter is an essential consideration, as TTS have surface gravities between those of dwarfs and giants.  Studies comparing these templates indicate that dwarf standard stars provide an acceptable match to TTS photospheres at optical wavelengths \citep{bb90}.  However, many of the strong NIR absorption features are gravity-sensitive, and the effect of surface gravity on the molecular bands near 1.65 $\mu$m produces a distinctive triangular shape in the continuum emission that must be accounted for when determining the excess over that region.\citep[see][Fig. 9.6]{gc09} 

%shape of the continuum and Additionally, between 1.3 and 2.0 $\mu$m, there are broad H$_2$O bands that increase in strength with lower surface gravity, producing a distinctive triangular shape in the continuum emission from 1.4 to 1.8 $\mu$m  \citep[see][Fig. 9.6]{gc09} that must be accounted for when determining the excess over that region.  

Here we use an infrared spectrograph, SpeX, to study the NIR excess in ten accreting TTS.  SpeX is the ideal instrument for this project, as it obtains continuous 0.8 to 2.5 $\mu$m coverage.  In this paper, we develop methods of veiling determination that account for the intermediate surface gravities of TTS and identify a set of lines that appear to have minimal effects from chromospheric activity.  Using these veiling values and WTTS template stars, we extract the excess above the photosphere in each of our targets.  We then perform a model-independent black body analysis to place constraints on the temperatures and emitting areas of the excess component, testing in the process the possibility of optically thick emission from inside the dust sublimation radius.

% with adequate spectral resolution and relative flux calibration in a single exposure, and it is possible to acquire near-simultaneous observations of the star with the guide camera to obtain an absolute flux calibration

\section{Observations and data reduction}
\label{obsred}

\subsection{Sample Selection of CTTS and WTTS}

Our initial sample of eight stars was selected from CTTS in the Taurus-Auriga molecular cloud complex for which we have \emph{Spitzer} Infrared Spectrograph (IRS) spectra \citep{furlan+06} that 1) do \emph{not} indicate the presence of gaps or other radial structure in their disks, 2) show a range of apparent excess in the NIR, based on 2MASS photometry and their spectral types in the literature, and 3) show a range of mass accretion rates from the literature.  Additional observations of two bright WTTS were included at our two most common spectral types (K7 and M2) for use as photospheric templates.  The CTTS are all single stars, within a detection limit of $\Delta m_K$=2 at separations greater than 20 mas, while the WTTS, LkCa 3 and V827 Tau, are sub-arcsecond binaries \citep{kraus+11}.  For the purposes of studying the excess over the photosphere, using binaries is obviously un-ideal.  However, based on their colors, the companions to LkCa 3 and V827 Tau are M6 and M5 and contribute less than half of the total flux in the NIR (45\% at $K_s$ for LkCa 3 and 37\% at $H$ for V827 Tau) \citep{wg01,kraus+11}.  Since these types will peak around $H$ or $K$ band \citep{kh95,rayner09}, they should be responsible for less flux than this at $i$, $z$, and $J$ bands.   Nonetheless, in \S\ref{ttsphot}, we demonstrate several techniques and checks to minimize the impact of the WTTS' binarity on our analysis.  The 2MASS magnitudes and stellar parameters of our sample are given in Tables \ref{sampletab} and \ref{littab}, respectively.  For each target we obtained spectra and photometry.

\subsection{Spectroscopy}
The spectra were obtained with SpeX \citep{ray03} at the NASA Infrared Telescope Facility (IRTF) on 1, 2, and 3 December 2010.  We observed our targets with the long-wavelength, cross-dispersed 2.1 $\mu$m mode (LXD) with the 0\farcs5$\times$15\farcs0 slit (R=$\lambda$/$\Delta\lambda$=2000) covering 2.1 to 4.5$\mu$m.  Following each LXD observation, we also observed the targets in the short-wavelength, cross-dispersed mode (SXD) with the 0\farcs3$\times$15\farcs0 slit (R=2000) from 0.8 to 2.5 microns.  Our SXD integration times were selected to have $S/N > 100$ in the $H$ band, while our LXD times were selected for $S/N\sim$20 at the $K$ band.  The data were obtained with the slit rotated to the parallactic angle and in an ABBA nod pattern.  

The spectra were reduced with the Spextool package \citep{cus04}.  We sky-subtracted each individual exposure using the opposite nod positions, extracted them separately, scaled each spectrum in the set to the collective median value, and combined them using the robust median option.  Spextool includes a package to correct for telluric absorption and perform relative flux calibration \citep{vac03} using observations of A0 stars.  These stars provide a relatively featureless continuum in the infrared against which the telluric absorption features can be clearly identified and removed.  Relative flux calibration is achieved by comparison of the telluric standard with a high-resolution model of Vega.  This process first requires the removal of the intrinsic hydrogen series absorption lines in the underlying spectrum of the telluric standard, which can be accomplished using either the instrument profile or a convolution between the telluric standard and Vega model.  The advantage of convolution is that it produces a better correction than the instrument profile; however, it requires an individual hydrogen line with sufficiently high $S/N$.  For these data, we were able to perform the convolution with Vega using the Paschen $\delta$ and Brackett $\gamma$ lines, respectively.  

Unfortunately, half of our observations in SXD had airmass differences greater than 0.1 between the targets and the nearest telluric standard observation.  All but two of the target observations were bracketed in airmass between two telluric observations.  For the targets with a poor match in airmass to any individual A0 star observation, we were able to produce an improved correction by taking an average of the two spectra of the program star, each corrected with one of the bracketing telluric observations and weighting by the airmass difference with the program star.

\begin{figure*}
\includegraphics[angle=0, scale=0.7]{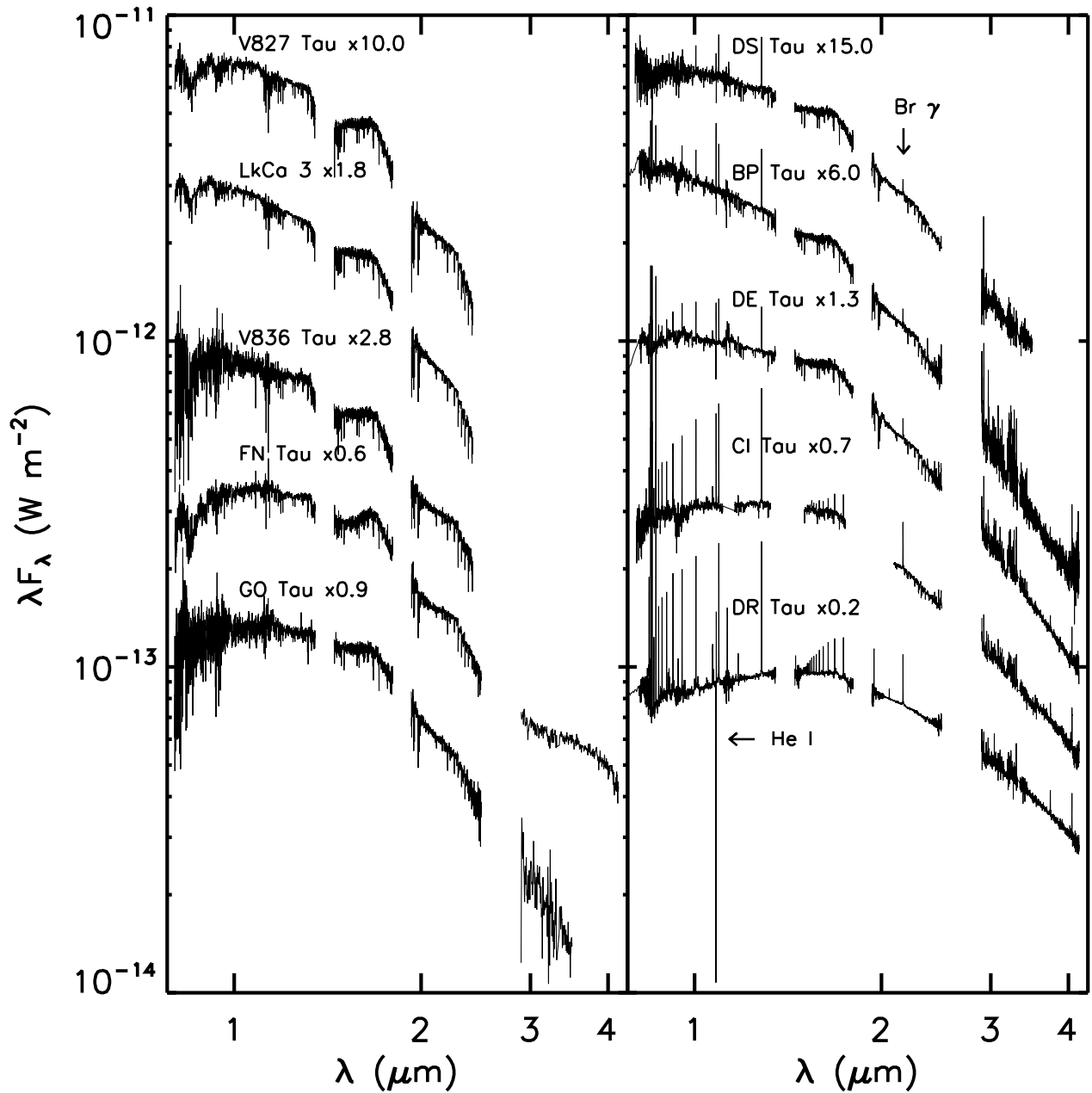}
\caption{Spectra of the ten stars in our sample.  {\it Left panel}:  The two WTTS stars (V827 Tau and LkCa 3) and the three low-accreting stars (V836 Tau, FN Tau, and GO Tau) from 0.8 to 4.5 $\mu$m.  The 2.8 to 4.5 $\mu$m regions of GO Tau and FN Tau, our faintest stars observed with LXD, have been smoothed to a lower resolution to increase their continuum signal-to-noise ratio.  {\it Right panel}:  The five higher accreting CTTS stars from 0.8 to 4.5 $\mu$m, in order of increasing Br$\gamma$ strength, i.e. accretion rate.  \label{rawwtts}}
\end{figure*}

Finally, the telluric-corrected, combined spectrum for a given order was merged with the neighboring order to produce a continuous spectrum.  We then excised the portions of each spectrum corresponding to the large telluric bands with less than 10\% transmission, e.g. $1.35<\lambda<1.45$, as these regions were too noisy to use.  The LXD spectra were scaled to the SXD spectra between 2.29 and 2.35 microns and spliced together at the point of equal $S/N$.  The final spectra are displayed in Figure \ref{rawwtts}.

\subsection{Photometry}
To flux calibrate our spectra, following each observation we imaged the target with the guide camera (0.1185 arcsec/pixel) at $K$ in a 7-position dither pattern.  We then observed a photometric standard from the extended list of UKIRT faint standards \citep{hawarden+01}, which was typically fainter than the targets.  Conditions were photometric, and integration times were chosen to maintain the same number of counts per pixel between the target and standard to avoid differences in the level of non-linearity.  The binary WTTS were not resolved, and the photometry listed for them in Table \ref{sampletab} is for both components.  The unregistered images of the photometric standard were median combined to create a master flat, as they typically had longer exposure times.  After applying the flat field image to the data for the target and standard, we registered and median-combined the flat-fielded images for each target.  Photometry was extracted from the final image using IRAF's \emph{phot} routine, with a gain of 14 $e^-/DN$, an instrumental zero-point magnitude of 20.57 at K, an aperture of 10 pixels, and a sky annulus from 30 to 35 pixels.  For each night we then derived a photometric solution using the observations of our standards at all airmasses and applied it to the data.  The resulting photometry is given in the last column of Table \ref{sampletab}; roughly half of the targets differ by at most 0.1 mag between our measurement and their 2MASS magnitudes, while the other half differ by up to 0.5 mag.  Since T Tauri stars are highly variable, these differences are likely genuine. 

\subsection{Ancillary Spectroscopy} 
\label{ancspec}
To test our infrared spectral typing, we needed optical spectral types obtained with a comparable technique, i.e. comparison of equivalent widths between the TTS and photometric standards.  To this end we used archival, low-dispersion optical spectra obtained at the 1.5m telescope of the Whipple Observatory with the Fast Spectrograph for the Tillinghast Telescope \citep[FAST;][]{fabricant+98} on the Loral 512$\times$2688 CCD.  Spectral coverage of $\sim$3600-7500 $\AA$ with a resolution of $\sim$6 $\AA$ was achieved with the standard configuration used for FAST COMBO projects: a 300 groove mm$^{-1}$ grating and a 3$^{''}$ wide slit.  The data were reduced at the Harvard-Smithsonian Center for Astrophysics using software developed specifically for FAST COMBO observations and were wavelength-calibrated and combined using standard IRAF routines. Spectra for all of our targets were obtained in 1995 and 1996 as part of Program 30 \citep[PI: Kenyon;][]{kenyon+98} and are publicly available in the FAST database\footnote {FAST database: http://tdc-www.harvard.edu/cgi-bin/arc/fsearch}.

\section{Analysis: Separating the Excesses and T Tauri Photospheres}
\label{ttsphot}

To constrain the physical parameters of the regions in which the NIR excess is emitted, we need to extract the flux-calibrated spectrum of this excess as a function of wavelength.  Since the excess fills in or `veils' absorption lines, we can measure it by comparing line strengths of the program stars with those of photospheric templates, in this case the two WTTS. To obtain the absolute flux of the excess, we also need the extinction towards the star, $A_V$, which can be estimated by comparing the slopes of the program star and the template.  However, the line strengths of the program star are affected by both veiling and the star's spectral type (SpT), while the slopes are affected by those parameters and $A_V$, so we need to determine SpTs, veiling, and $A_V$ simultaneously from the same data set.

The degree of continuum veiling can be defined by the term $r_{\lambda}=\frac{F_V(\lambda)}{F_c(\lambda)}$ and related to the line, continuum, and excess fluxes, $F_l$, $F_c$, and $F_V$, respectively, by:  \begin{equation}
\label{hegmeth}
\frac{F_l(\lambda)}{F_c(\lambda)} = \frac{F^{phot}_l(\lambda)+F_V(\lambda)}{F^{phot}_c(\lambda)+F_V(\lambda)}=\frac{\frac{F^{phot}_l(\lambda)}{F^{phot}_c(\lambda)}+r_{\lambda}}{1+r_{\lambda}}
\end{equation}
\noindent where the superscript $phot$ indicates the intrinsic flux of the underlying photosphere.  \citet{heg95} use Eq. \ref{hegmeth} to fit a single value of $r_{\lambda}$ to several lines at once, and this has become the most common technique used to determine veiling in the recent literature \citep[e.g.][]{muzerolle03,fischer+11}.  Alternatively, the veiling can be written in terms of the equivalent width, $W_{\lambda}$, assuming that the veiling is constant over the line:
\begin{equation}
\label{ewveil}
\frac{W^{phot}(\lambda)}{W(\lambda)}=1+r_{\lambda}
\end{equation}
\noindent \citet{bb90} compare the two approaches and find that veilings determined by the latter method have less scatter, as one can reject lines known to experience differential veiling or surface gravity effects.  We therefore use the equivalent width method to determine the veiling for each of our targets.

Equivalent widths of particular atomic lines and molecular bands are also used to determine SpTs, e.g. \citet{hernandez+04}.  The effects of continuum veiling on the SpT determination can be eliminated by considering instead the \emph{ratio} of equivalent widths of two lines close enough in wavelength that their veiling is approximately constant \citep{bb90,lr98,vacca11}.  Then their ratio should be equal to that of a photospheric template of the appropriate spectral type.  What constitutes ``close enough'' is unclear.    \citet{hartigan+89} find the veiling to be constant over 10 to 15{\AA} intervals at optical wavelength, where the veiling continuum is increasing in strength towards the UV.  The slope of $r_{\lambda}$ vs. $\lambda$ becomes flatter in the infrared \citep{wh04}, so the veiling is likely constant over a larger interval at the SpeX wavelengths.

\subsection{Continuum Determination and Equivalent Widths}

\begin{figure*}
\includegraphics[angle=0, scale=0.6]{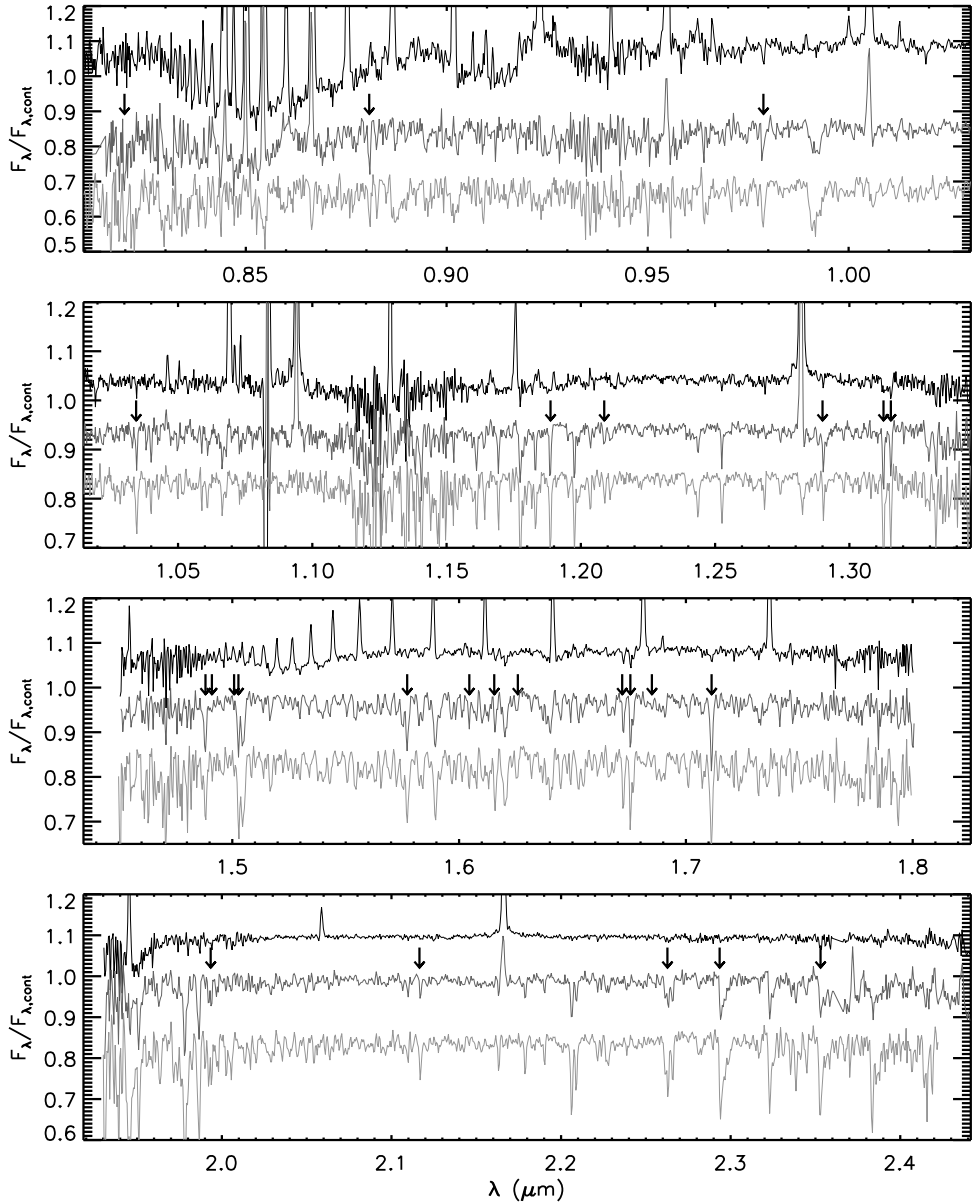}
\caption{Continuum normalized SXD spectra for DR Tau (black), the strongest accreter in our sample, DE Tau (dark grey), a moderate accreter, and LkCa 3 (light grey), our M2 WTTS over the following bandpasses: {\it (top)} $i$ and $z$ bands, {\it (middle-top)} $z$ and $J$ bands, {\it (middle bottom)} $H$ band, and {\it (bottom)} $K$ band. Of these regions, $H$ band typically has the best S/N.  The absorption features used in this analysis are indicated with arrows, with identifications given in Table \ref{veilings}.  We note that practically all of the absorption features in these spectra are real, spectrally unresolved blends of lines and refer interested readers to \citet{rayner09} for more detailed information.  \label{allewfits}}
\end{figure*}

The first step in the procedure outlined above was to measure the equivalent widths of individual lines.  This is relatively straightforward in certain spectral regions, e.g. $J$ or $K$ band, but in $z$ and $H$ band there are many molecular absorption features overlapping the atomic lines. Consequently, even if one does find a ``good'' absorption feature to measure, i.e. one with mainly one contributing absorber and good signal to noise, it is still difficult to define continuum regions for extracting the equivalent width.  To overcome this issue, we use a technique outlined by \citet{bb90} for identifying the continuum wavelengths.  To summarize, we choose as continuum points the wavelengths corresponding to the highest fluxes in the distribution of fluxes within some wavelength bin, after performing a $\sigma$-clip to remove obvious emission lines.  We fit these points with a polynomial, divide out the continuum, fit each absorption line with a gaussian, and use the analytic expression for the area of the gaussian to compute the equivalent width.  Uncertainties are propagated from the original uncertainties in the flux, and we employ an additional criterion that the equivalent width of the line must be greater than 0.2 {\AA}, which is the $rms$ uncertainty in regions of the spectra without obvious absorption lines.  The lines which we ultimately decided to use are shown in the continuum subtracted and normalized spectra of one of our WTTS and two CTTS in Fig. \ref{allewfits}.  Identifications for each line are given in Table \ref{veilings}.

\subsection{Comparison of Trends in $W_{\lambda}$ vs SpT Between WTTS, Dwarfs, and Giants}
\label{trendcomp}

We began our analysis with two assumptions.  First, a WTTS with no infrared excess should be a representative photospheric template for a CTTS of the same effective temperature; i.e. the WTTS has no intrinsic continuum veiling.  Second, we assumed the \citet{bb90} result that there are optical lines for which dwarf standard stars are acceptable representations of WTTS effective temperatures, and that if a star is spectral typed with these lines, any deviations from the standard dwarf trends between $W_{\lambda}$ and SpT from 0.8 to 2.5 $\mu$m are due to the effects of surface gravity or differential veiling caused by chromospheric emission, as discussed in \S \ref{intro}.  In this way, we can use our WTTS standards to identify which infrared lines are surface gravity sensitive in TTS.  Because our particular WTTS are binaries, we operated with the additional caveat that the lines we ultimately chose needed to be more sensitive to spectral types in the range of the primaries, K7 to M2, than to late M types, i.e. the spectral types of their companions.

To confirm that our WTTS standards had no excess and to determine qualitatively which of the CTTS were the most veiled,  we constructed equivalent width vs spectral type plots for the IRTF Spectral Library \citep{rayner09} dwarf and giant standard stars over a sample of lines, as described in Appendix \ref{app2}.  Since some `lines' were obviously spectrally unresolved line blends, not all of the features produced a clear trend as a function of spectral type. However, there were several lines for which the equivalent width depended strongly on the spectral type over the range predicted for our targets, $\sim$ K5 to M5.  

We then derived optical spectral types for our TTS sample from the ancillary optical spectra (described in \S \ref{ancspec}) with the SPTCLASS tool\footnote {SPTCLASS code: http://www.astro.lsa.umich.edu/~hernandj/SPTclass/sptclass.html}, an IRAF/IDL code based on the methods described by \citet{hernandez+04}. The code computes spectral types for low mass stars (K to M5) by measuring the equivalent widths of 16 spectral features that are sensitive to changes in the stellar effective temperature.  Each spectral index is calibrated using spectroscopic standards observed with FAST.  For most of the program stars, these spectral types agree with those from the literature to within 0.5 subclasses.  We then overplotted the $W_{\lambda}$ for our sample on the dwarf and giant $W_{\lambda}$ trend plots, assuming the TTS optical spectral types from SPTCLASS.  

From this comparison, there are lines over our whole wavelength range for which the WTTS lie on the dwarf trend curve at locations consistent with their optical spectral type, and in these lines it is easy to see the degree of veiling in the CTTS relative to the WTTS (see Al I doublet at 1.31270 $\mu$m in Appendix \ref{app2}, Fig. \ref{gravtrendsa} for example).  In addition to our two WTTS, we identified two CTTS that have very low accretion rates and therefore are not veiled relative to the dwarf standards at $z$, $J$, $H$, or $K$ bands at the resolution of our observations: FN Tau and V836 Tau.  Both have infrared excesses from 5 to 40$\mu$m indicative of dusty disks, as seen by {\it Spitzer} \citep{furlan+06}.  Because these CTTS are single within the limits of our selection criteria \citep[see \S\ref{obsred},][]{kraus+11} and have little to no veiling at our spectral resolution and sensitivity, we can use them as a check on the position of the WTTS in $W_{\lambda}$ vs. SpT diagrams to ensure that we are not getting contamination from the WTTS' companions in our $W_{\lambda}$ measurements.   We will refer to these two stars henceforth as `weakly veiled CTTS' and assume that their SPTCLASS optical spectral types are accurate, as we did with the WTTS.  

Using the two WTTS and the two weakly veiled CTTS we define rudimentary $W_{\lambda}$ vs. SpT trends for each line.  It is obvious from the trends that the five other TTS are veiled, and that for most of the lines, this veiling is degenerate with the spectral classification.  That is to say, a K7 star with moderate veiling could have the same equivalent width as either a K3 or M3 star with little veiling, depending on the shape of the trend.  In addition, almost all of the deep lines ($> 1{\AA}$) are surface gravity sensitive, as indicated by the discrepancies between the dwarf and giant standard trends.  This effect is particularly noticeable for lines in the $H$ band (e.g. Mg I 1.57721 $\mu$m line in Appendix \ref{app2} Fig. \ref{gravtrendsa}).  In general, the TTS trends fall between the dwarf and giant trends, although the degree to which this is true varies on a line-by-line basis.  

There are some lines, e.g. Mg I at 1.1833 $\micron$, that show shallower features than can be accounted for by surface gravity effects (Appendix \ref{app2}, Fig. \ref{gravtrendsa}).  For these lines, even the WTTS appear to be veiled.  The effect is that for these lines, even the WTTS appear to have a later infrared spectral type than their optical spectral types.  This may be an effect of differential veiling.  Furthermore, for some of the shorter wavelength lines that peak in $W_{\lambda}$ at mid-M types, e.g. the 1.14 $\mu$m Na doublet (Appendix \ref{app2}, Fig. \ref{gravtrendsa}), the later three WTTS/weakly-veiled CTTS stars lie on the dwarf trend, but our $\sim$K7 WTTS, V827 Tau, and weakly-veiled CTTS, V836 Tau, have $W_{\lambda}$ that are much greater than the K7 point in the dwarf trend, making them appear to be later, $\sim$M1 to M2.  In the case of V827 Tau, the binary companion may be affecting this line, but for V836 Tau this may be caused by starspots, as it seems to affect some stars more than others of the same spectral type. 

To fit the TTS trends, we made both a linear fit to the WTTS and weakly-veiled TTS and a reduced $\chi^2$ based interpolation between the dwarf and giant trends for each line.  These fits for a set of typical lines are displayed in Appendix \ref{app2}, along with the fraction of the curve attributed to the dwarf trend vs. the giant trend. 

\subsection{Spectral Types}
\label{sptsection}

Since SPTCLASS does not account for continuum veiling, we computed infrared spectral types for each of the veiled CTTS using ratios of equivalent widths of two nearby lines to avoid the effects of continuum veiling in the measurement of the spectral types.  Because the lines for which there was no surface gravity dependence were typically not `close enough' in wavelength to use those exclusively for such ratios, we ultimately used a combination of those lines and lines for which the surface gravity effects were well-fit by our TTS trend. The details of our line selection are given in Appendix \ref{app3}, and in Fig. \ref{sptrat}, we display the line ratios that we ultimately used.  The resulting spectral types for our sample are given in Table \ref{starparams}.

Comparing the derived infrared spectral types with those from the literature and from SPTCLASS, listed in Table \ref{littab} and \ref{starparams}, the SPTCLASS SpTs are in agreement with the \citet{kh95} spectral types, within the error bars, for all of our program stars except for FN Tau (M3 vs M5) and DR Tau (K3 vs K7).  FN Tau is highly variable and known to flare, while DR Tau is highly veiled, which would make strong metal lines looker weaker and earlier.  This is likely why our veiling-independent infrared spectral type for DR Tau is later (K7-M0).  Aside from those stars, our infrared spectral types are also consistent within the uncertainties with the optical SpT from SPTCLASS and \citet{kh95}.  This suggests that the differences between optical and infrared spectral classifications in the literature may be the result of intrinsic differences between TTS, dwarf, and giant standard stars, such as surface gravity, that are stronger at infrared than optical wavelengths.

\subsection{Veiling}
\label{regveiling}

\begin{figure*}
\includegraphics[angle=0, scale=0.9]{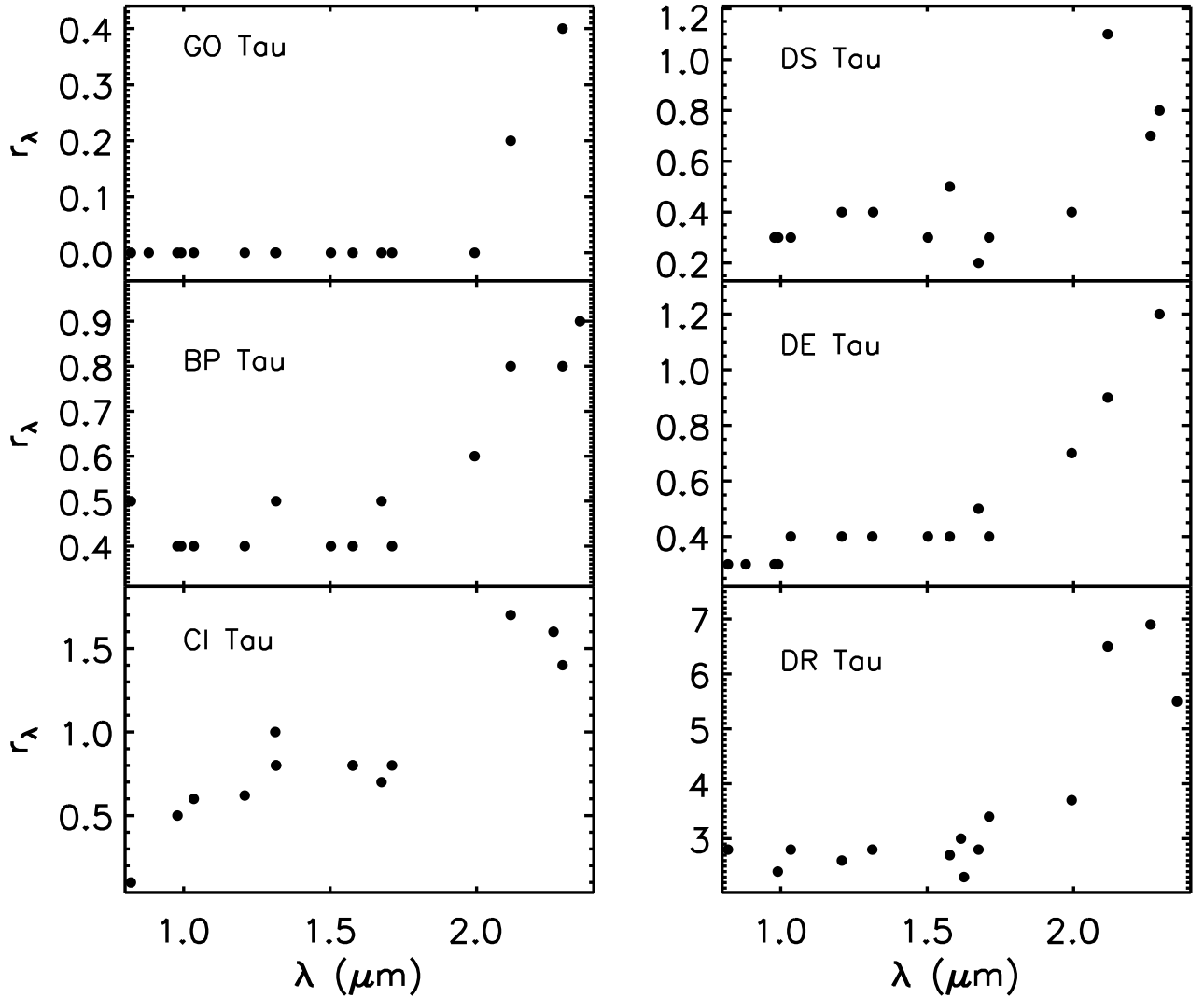}
\caption{Veiling, $r_{\lambda}$, as a function of wavelength for the final sample of lines and the six veiled CTTS. Systematic uncertainties are estimated as $\sim$0.2 based on the uncertainties in the TTS trends. We note that the actual uncertainties for the more veiled targets, CI Tau and DR Tau, are likely larger ($\sim$ 0.5-1.0) and mostly dependent on how the continuum is defined.  \label{rlambdanofe}}
\end{figure*}

Having established the spectral types of our program stars, we now use these spectral types to determine veilings over the 0.8 to 2.3 $\mu$m range.  Taking the surface-gravity interpolated TTS trends to represent the intrinsic equivalent widths for the TTS in each line, we computed the veilings, $r_{\lambda}$, from Equation (\ref{ewveil}).  Initially we included all of the lines for which we could determine a TTS trend.  However, this resulted in a large dispersion in the veiling of nearby lines, particularly at $J$ and $H$ band.  To explore this, we compared these lines across the range of CTTS spectral types.  Many of the features which produced exceptionally large or small values of $r_{\lambda}$ at $z$, $J$, and $H$ in the moderately veiled stars were close to lines which, in the more highly veiled stars, went into emission.  For the more moderately veiled stars, these emission lines were sometimes too weak to be picked up by our $\sigma$-clipping threshold, which had the effect of both filling in some lines, making them appear more veiled, or skewing the continuum fits, which made them appear either more or less veiled, depending on if the continuum fit too high or too low.  To produce a sample of veilings that reflected the continuum veiling only, we excluded all features at wavelengths near emission lines appearing in CI Tau and DR Tau and refined our continuum fits for some of the 0.8 to 1.0 $\mu$m  lines to have a longer baseline.  The final veilings for each wavelength of the sample are presented in Fig. \ref{rlambdanofe}.  The dispersion is significantly less at $z$ band, although in $H$ band there is still some scatter.  We list these final veilings in Table \ref{veilings}.

\subsection{Extinction, Stellar Parameters, and Excesses}
\label{avsandexcesses}

To estimate the amount of extinction along the line of sight towards each of our program stars, we use the relationship between the observed target fluxes, the (extinction corrected) photospheric template fluxes, and the veiling of the photospheric template at a given wavelength.  The degree of veiling should be the same before and after extinction correction, so some minor arithmetic and the assumption of an extinction curve, $A_{\lambda}/A_V$  leads to:

\begin{equation}
\label{alam}
A_{V}=\frac{1}{A_{\lambda}/A_V}2.5\,{\rm log}\left((1+r_{\lambda})\frac{F_{phot}(\lambda)}{F_{t,obs}(\lambda)}\right)
\end{equation}

\noindent This relationship can be used to compute $A_V$ from a linear fit with $A_{\lambda}/{A_V}$ as the abscissa, $2.5\,{\rm log}((1+r_{\lambda})F_{phot}(\lambda)/F_{t,obs}(\lambda))$ as the ordinate, and $A_V$ as the slope \citep{gullbring+98a,sd+08,fischer+11}.

\begin{figure}
\includegraphics[angle=0, scale=0.65]{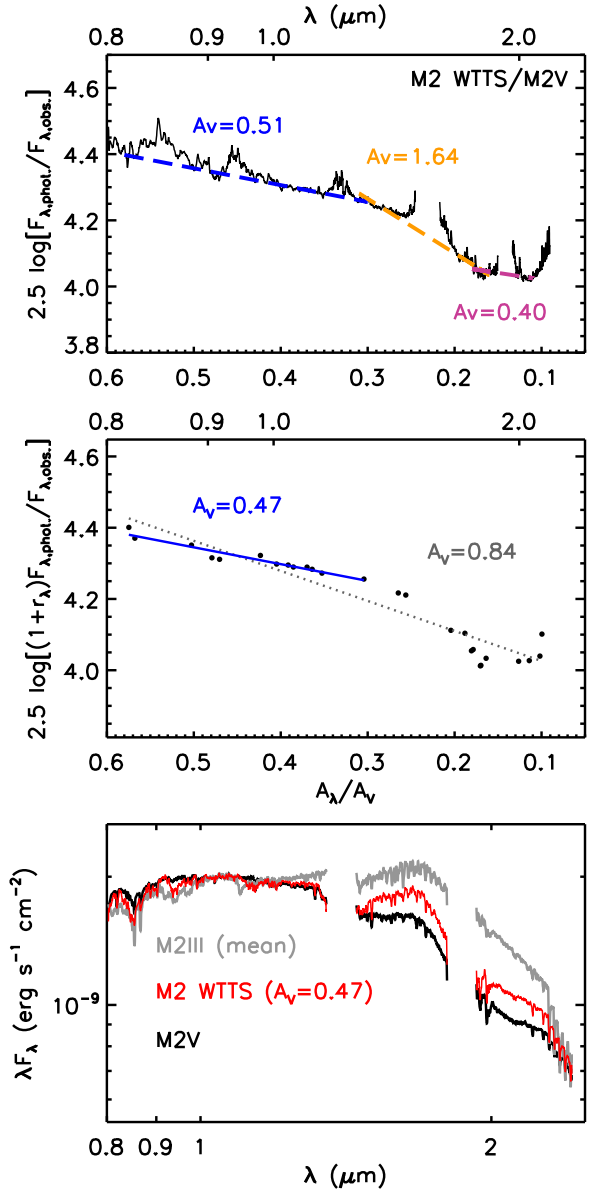}
\caption{{\it (Top panel)} Comparison between shape of LkCa 3 and an M2V standard.  $A_V$ is the slope, assuming no veiling.  The slope changes around $A_{\lambda}/A_V$=0.3 and again around $A_{\lambda}/A_V$=0.18.  {\it (Middle panel)} Plot of the relation given by Equation (\ref{alam}) for LkCa 3 ($r_{\lambda}=0$).  $A_V$ determined from the two linear fits to the blue or entire wavelengths ranges are listed at top.  {\it (Bottom panel)} Surface gravity dependence of continuum shape:  observed LkCa 3 spectrum (red), extinction corrected, and the M2V standard (black) and M2 III standard (light grey), scaled to the observed spectrum at 1.1 $\mu$m.   The M2III spectrum is the average of the IRTF library M1III and M3III, as the nominal M2III spectrum in the library appears closer to an M4 III in shape than an M2 III.  Note that although the continuum shape of the WTTS appears `later' than the dwarf standard of the same spectral type, it lies between the dwarf and giant standards and can be better explained by a difference in surface gravity.  \label{bptauav}}
\end{figure}

To test how much our $A_V$ determination would be affected by choosing either one of our binary WTTS or a dwarf standard as the photospheric template, we construct such a plot for our M2 WTTS, LkCa 3.  As seen in the top panel of Fig.  \ref{bptauav}, there is a clear change in slope (and therefore the implied $A_V$) between the 0.6$\ge A_{\lambda}/A_V \ge$0.3 region and the 0.3$\ge A_{\lambda}/A_V \ge$ 0.15 region.  In the middle panel of Fig.  \ref{bptauav}, we demonstrate how this change in slope affects the determination of $A_V$.  For LkCa 3, fitting a line to only the data with $A_{\lambda}/A_V \ge$0.3 (the $z$ and $y$ bands) produces $A_V$=0.45, which is what is found for optical determinations for this star.  However, if one fits a line over the whole range of $A_{\lambda}$/$A_V$, the resulting slope is steeper than that of the $z$-$y$ band $A_V$ by almost a factor of 2, meaning that we have overestimated $A_V$.  Extinction correcting the LkCa 3 spectrum to 0.45 mag and the M2V standard to it reveals the source of the discrepant $A_V$:  there is an intrinsic difference in the shape of LkCa3 and M2V standard at H and K bands (bottom panel, Fig. \ref{bptauav}).  This difference is greater than what can be attributed to a change in slope induced by the LkCa 3 companion and disappears by 2.5$\mu$m.  In contrast, the shape of the excess above the dwarf standard photosphere is consistent with the lower surface gravity seen in TTS; the WTTS lies between the dwarf and giant M2 standards for all wavelengths between 1.4 and 2.5 $\mu$m.  

\begin{figure}
\includegraphics[angle=0, scale=0.5]{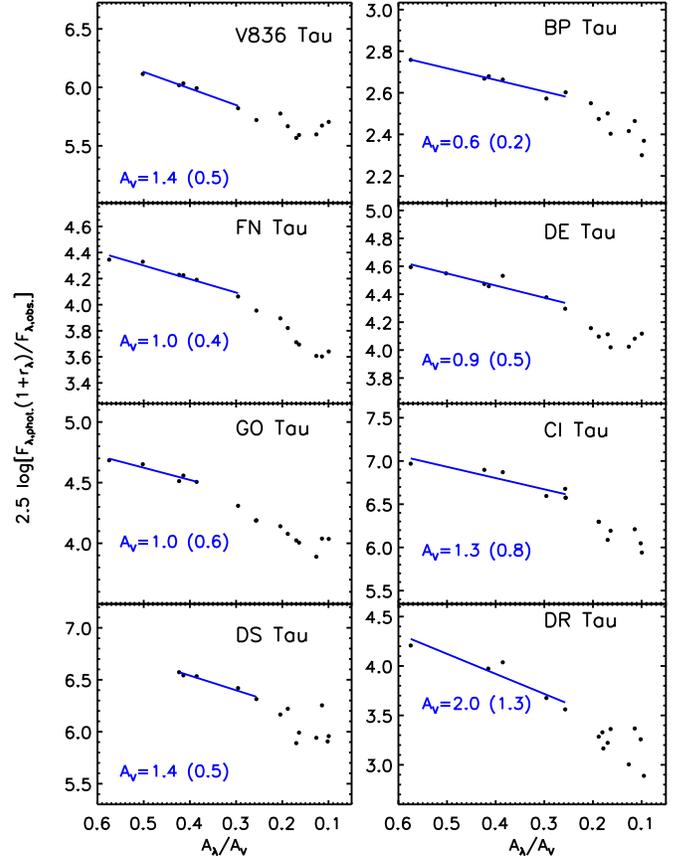}
\caption{$A_V$ determination for the entire sample using Equation (\ref{alam}) and dwarf standard stars.  $A_V$ is the slope of the linear fit.  Fits to the whole wavelength range may be skewed by surface gravity effects, as demonstrated in Fig. \ref{bptauav}.  We chose to fit the region least affected by surface gravity (0.8 to 1.2 $\mu$m; blue line) to obtain our final $A_V$s, which are listed in Table \ref{starparams}.   \label{allavs}}
\end{figure}

Given that the dwarf standard continuum shape matches the WTTS outside of TiO bands from 0.8 to 1.3 $\mu$m and that the veilings we measured should be least influenced by the binarity of our WTTS over those wavelengths, we fit $A_V$ over that region, only, for our other program stars.  To obtain values for $A_{\lambda}/A_V$ over our wavelength range we fit a splined curve to the $R_V$ = 3.1 extinction law of \citet{mathis90} and interpolate from that curve to the wavelength of each veiling estimate.  Three-sigma uncertainties in $A_V$ are determined from the uncertainty in the slope parameter of the linear fitting routine and are between 10 and 65$\%$ of the total values.  Excesses above the photosphere are constructed by extinction correcting the program star spectra using the derived $A_V$, scaling each dwarf photospheric template by the average veiling at $\sim$1 to 1.3 $\mu$m (where the veiling is the smallest), and subtracting the scaled templates from the TTS. The $A_V$ determinations are plotted in Fig. \ref{allavs}, while the values of $A_V$, with uncertainties, are listed in Table \ref{starparams}, and the excesses for the sample are plotted in Figs. \ref{allbbfitsa} and \ref{allbbfitsb} as part of the analysis in \S \ref{bbfits}. 

After determining the best photospheric template, the position of the photosphere with respect to the observed spectrum, and the absolute, extinction-corrected flux of the photosphere for each our program stars, we compute their luminosities and radii.  We use the extinction corrected, scaled flux of the photospheric template at $J$ to obtain the absolute $J$ magnitude.   Assuming the spectral types found earlier, which are consistent with dwarf spectral types at optical wavelengths, we use the effective temperatures and colors in Table A5 with Equation (A1) in \citet{kh95} to derive the absolute bolometric magnitude for each star, and from there the luminosity and radius.  Our results are listed in columns 4 and 5 of Table \ref{starparams}.  Finally, we assume the PMS evolutionary tracks of \citet{siess2000} to determine the mass of each program star and record those values in the same table.  It should be noted that although evolutionary tracks are often used to determine mass, this method is highly uncertain and the results dependent on the tracks assumed. However, the uncertainty is systematic, so we do not propagate it here.

Comparing our derived values of $A_V$ and $L_*$ with those from the literature (see Table \ref{littab}), we note that within our uncertainties, we are consistent with previous estimates in most cases.  Our values of $A_V$ are not systematically higher or lower than those previously found.  Comparing the two targets which we have in common with \citet{fischer+11}, BP Tau and DR Tau, our $A_V$ for BP Tau is lower than their value (0.6 vs 1.75 mag), while our $A_V$ for DR Tau is larger (2.0 vs 1.54 mag).  For DR Tau, their $A_V$ is within the uncertainty of our value, and for BP Tau, the difference is clearly the result of fitting to different wavelength regions; as seen in their Fig. 7, determining $A_V$ over $A_{\lambda}/A_V$ from 0.6 to 0.3 only would produce a lower value.  

\section{Analysis: Parametric Fits to the Excess}

Having derived the relevant stellar parameters, extinction along the line of sight, and NIR emission excesses we turn our attention to disentangling the contributions to the emission excess.  From the excess plots constructed in \S \ref{avsandexcesses}, we see that the five most veiled CTTS, DS Tau, BP Tau, DE Tau, CI Tau, and DR Tau, have an emission excess over the entire 0.8 to $\sim$5 $\mu$m range, while our three weakly veiled CTTS, V836 Tau, FN Tau, and GO Tau have no excess at the shorter wavelength end, but some small excess from 2 to 3 $\mu$m.  As discussed in the introduction, the NIR excess originates primarily in a dust sublimation wall, with potential contributions from the longer wavelength end of emission from an accretion shock or optically thick gas inside the dust sublimation radius. We leave such detailed physical modeling to the second paper in this series. Here we opt to derive basic quantities such as the characteristic temperatures and solid angles for different regions of the excess, which is possible due to our absolute flux calibration and the wide wavelength coverage of the data.  The results of this analysis will be used to guide detailed modeling efforts in our second paper (Paper II).

\subsection{Characteristic Temperatures and Solid Angles}
\label{bbfits}

Because we expect to see the longer wavelength tail of blackbody-like emission from an accretion shock, plus blackbody-like emission from a dust sublimation wall, and a possible third emission component from optically thick gas, we decided to fit the excess with a simple parametric model consisting of three black bodies at hot, warm, and cool temperatures.  We were able to achieve a set of acceptable fits to the excesses by allowing all three temperatures to vary within specified ranges in a grid and optimizing the solid angle of each blackbody to best fit the combined sum of the blackbodies to the excess, identifying the fit that has the lowest value for $\chi^2$ per degree of freedom.  

\begin{figure*}
\includegraphics[angle=0, scale=0.7]{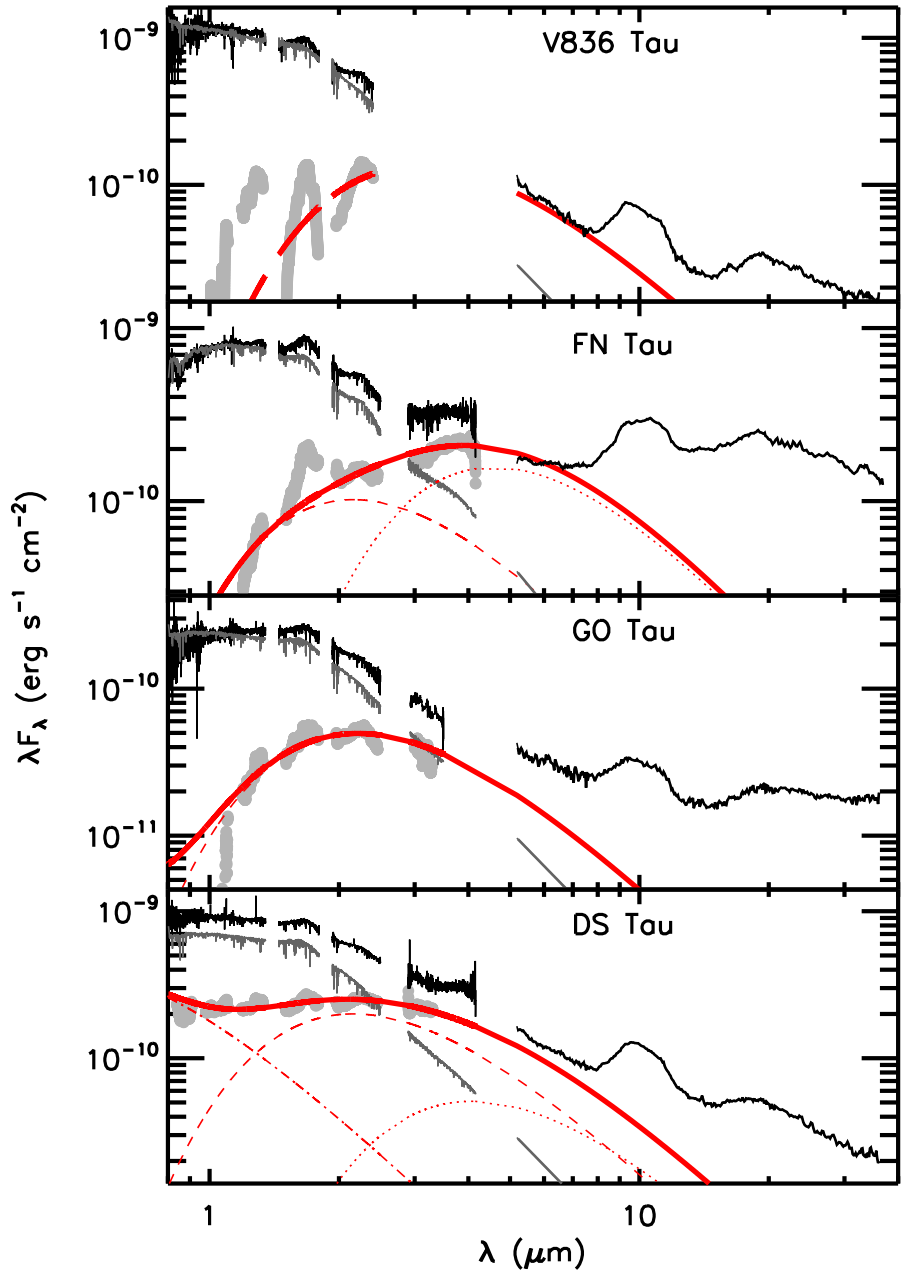}
\caption{Parametric fits to the emission excess for V836 Tau, FN Tau, GO Tau, and DS Tau.  The observed, extinction-corrected TTS spectrum, including both SpeX and Spitzer IRS from \citet{furlan+06} (black) and WTTS photospheric template (dark grey) are plotted as well for reference.  The excess above the photosphere (light grey, thick) is rebinned to a lower resolution for display purposes and fit by three blackbodies (red) with $T_{hot}$ (dashed-dotted), $T_{warm}$ (dashed), and $T_{cool}$ (dotted).  The combined fit is given by the solid, red line.  Values for the temperatures and solid angles are given in Table \ref{results}. We note that the IRS spectra were not included in the fit, but rather plotted for independent comparison. \label{allbbfitsa}}
\end{figure*}

\begin{figure*}
\includegraphics[angle=0, scale=0.7]{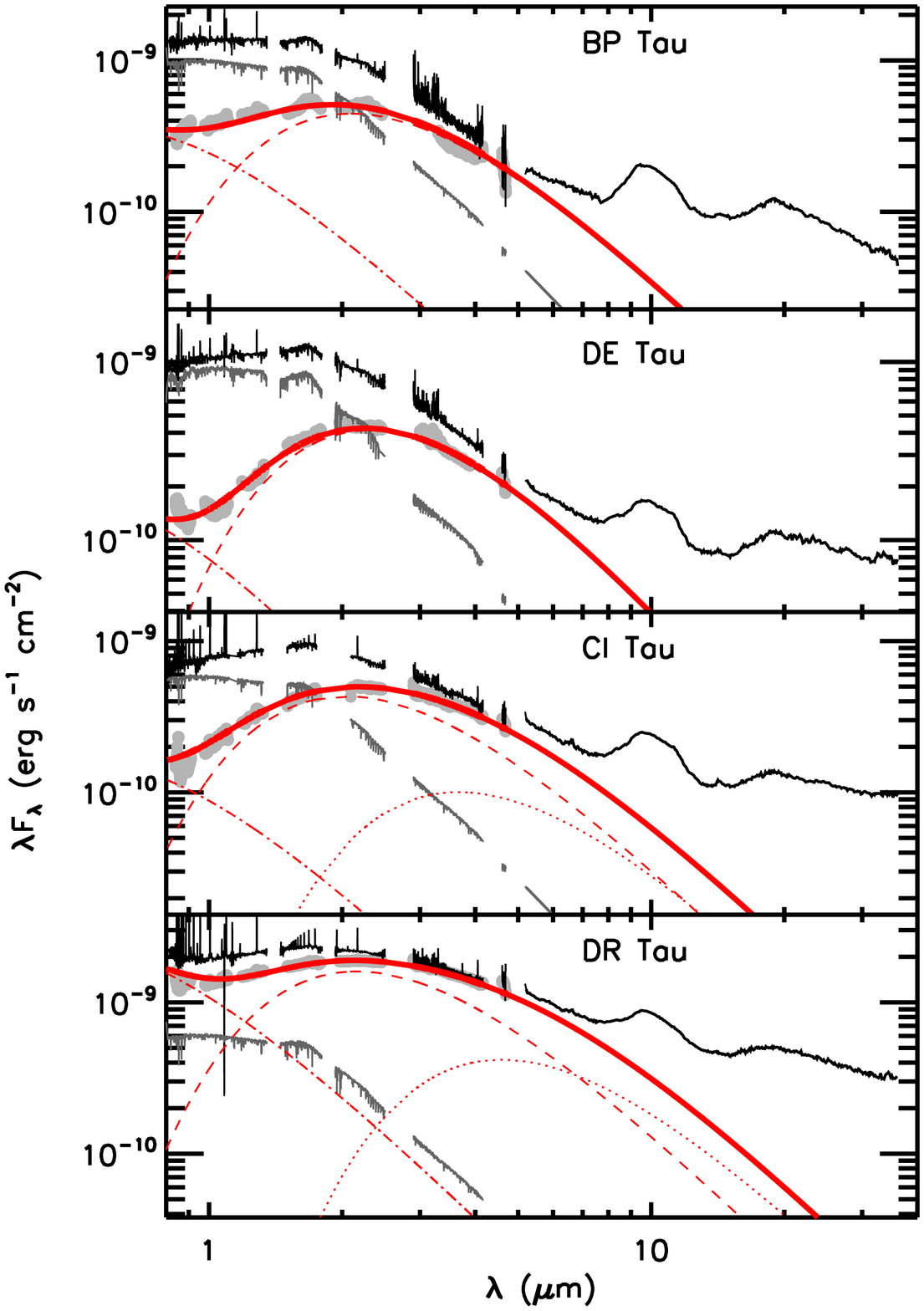}
\caption{Parametric fits to the emission excess for BP Tau, DE Tau, CI Tau, and DR Tau.  The meaning of each component is described in the caption for Figure \ref{allbbfitsa}.   \label{allbbfitsb}}
\end{figure*}

The cool temperature component was allowed to vary from 500 to 1500 K, with the intention of representing a typical dust sublimation temperature.  The warm temperature component ranged from 1500 K to the effective temperature of the star, intended to simulate the unidentified $H$ band excess (either from more highly refractory sublimating material or hot, optically thick gas just inside the typical dust sublimation region).  Both the cool and warm components varied in increments of 100 K.  The hot temperature component ranged from 4000 K up to 8,000 K in increments of 1000 K, and was intended to represent an accretion shock component \citep{cg98}.  For temperatures higher than $\sim$8,000K, our wavelength range only covers the Rayleigh-Jeans tail of the Planck function, so to first order the temperature of such a fit is degenerate with the solid angle.  Therefore when considering the quality of the fit, we also required that the flux produced by the best-fitting temperature and solid angle not exceed the observed $V$ band flux listed by \citet{kh95}, to reduce this degeneracy .  Although these stars are variable at $V$ band, the error in $m_V$ is between 0.15 and at most 0.4 magnitudes \citep{herbst+94}, much less than the range of error introduced by $A_V$ uncertainties (0.2 to 1.3 magnitudes).  The resulting fits are displayed in Figs. \ref{allbbfitsa} and  \ref{allbbfitsb} with the temperatures and solid angles tabulated in Table \ref{results}.

We found that none of the TTS were well fit with temperatures of 1400 K, typically taken as the dust-sublimation temperature.  Instead, they required temperatures consistent with what is expected for the dust sublimation temperature of high density, highly refractory grains; that is to say in the range of 1600 to 2000 K \citep{hemley94,posch+07}, which is consistent with the temperatures of the wall in Herbig AeBe systems \citep{mmg02}.  An additional \emph{cool} component is required to fit the excesses of DR Tau, CI Tau, DS Tau, and FN Tau.  The cool component has a temperature in the range of 800 K to 1000 K for those stars.  There are two other groups of stars: V836 Tau and FN Tau, which are best fit with a cool component and no warm component, and BP Tau, DE Tau, and GO Tau, which are best fit by a warm component only; i.e. their walls are well-represented by a single-temperature blackbody at $\sim$1700 K.  In all cases, the solid angles of the warm and cool components were much larger than the solid angle of the central stars.  The hot component was the only blackbody whose solid angle was less than that of the central star, which is consistent with it arising in the accretion shock.  The weakly veiled CTTS had so little veiling shortward of 1 $\mu$m that the `excess' in that region is completely noise.  Consequently, we do not list a temperature or solid angle for their hot components.  The more veiled stars had a range of $T_{hot}$ from 6,000 K to 8,000 K,  consistent with the temperature range of shocks given by \citet{cg98}.  

Since the CTTS DR Tau is also in the sample of \citet{fischer+11}, who fit its excesses with three black bodies and conclude that has an $H$ band excess that cannot be explained by the shock or wall component, we chose to see how well our excesses could be fit using their Case A and B temperature sets.  In both cases, their hot and cool components are fixed at 8000 K and 1400 K, respectively, which are intended to reflect the temperature of a hot shock component and the dust sublimation wall.  For Case A, the warm component is fixed at 5000 K, intended to represent a lower energy shock, while in Case B it is fixed at 2500 K, intended to represent optically thick gas in the inner disk.  The temperature sets in both Case A and Case B produced poorer fits to the extracted excess than our best-fitting values, with $\chi^2$/d.o.f. of 74.2 and 98.3, respectively, compared with a $\chi^2$/d.o.f. of 10.7 for our fit.  Therefore, while we cannot completely rule out the possibility that there is a contribution from optically thick gas in the inner disk of DR Tau, we do not require it to fit our excess.

\subsection{Accretion and Wall Luminosities}
\label{brgamma}

While our spectra do not cover the wavelength region over which the excess from an accretion shock would be strongest, we can make use of the Br$\gamma$ emission line at 2.17$\mu$m to estimate the accretion luminosity.  By integrating over the continuum subtracted, flux calibrated Br$\gamma$ line from the excess spectrum, we are able to obtain an estimate of the luminosity in Br$\gamma$.  Then from the correlation in Equation (2) of \citet{muzerolle98}, 

\begin{equation}
log(L_{acc}/L_{\odot})=(1.26\pm0.19)log(L_{Br\gamma}/L_{\odot})+(4.43\pm0.79), 
\end{equation}

\noindent we use the line luminosity to determine the accretion luminosity, $L_{acc}$, listed in Table \ref{excesslums}.  Br$\gamma$ emission was not detected in our two lowest-veiling TTS, V836 Tau and FN Tau.  GO Tau had a minimal detection.  Of the five more heavily veiled TTS, DS Tau, BP Tau, and DE Tau have accretion luminosities between 10 and 20\% of their stellar luminosities, while CI Tau and DR Tau are accreting with 90\% and 260\% of their stellar luminosities, respectively.  Comparing the values of $L_{acc}$ found here with those from the literature, given in Table \ref{littab}, we find that three of our targets, GO Tau, DS Tau, and BP Tau, all have lower $L_{acc}$ by 40 to 60\% in our analysis compared with previous estimates.  In contrast, DE Tau, CI Tau, and DR Tau all have higher $L_{acc}$ by 25 to 100\% in this work.  Some of the variation is certainly real; for example, GO Tau had no Br$\gamma$ emission above the noise in our spectrum, but clearly has $L_{acc}$ estimates in the literature.  A portion of the variation in our sample could be caused by the differences in our procedure for estimating $L_{Br\gamma}$.  Since we have flux-calibrated, extinction-corrected  excesses, we determine $L_{Br\gamma}$ directly rather than using the equivalent width and an estimate of the continuum from photometry from a non-simultaneous observation.  Ultimately, though, the biggest difference is likely in our estimates of $A_V$ via its effect on $L_*$.  From $L_{acc}$, we can calculate the mass accretion rates onto the stars, \Mdot, as \Mdot$=L_{acc}R_*/(GM_*)$, listed in the fourth column of Table \ref{excesslums}.  With the exception of DS Tau, all of the moderate to strong accreters have accretion rates greater than 10$^{-8}$\Msun$/yr$, the CTTS average value at 1 to 2 Myr \citep{gullbring+98a,wg01}.  We can also use these independent estimates of $L_{acc}$ to compare with the luminosities implied by our parametric fits to the excess and to determine the radius of the wall, below and in \S \ref{discussion}.  

As a brief check on the credibility of our fits, we compare the luminosities of the fitted blackbody components with those of the central star and the accretion luminosity derived from Br$\gamma$.  For each component, we determined the luminosity from the temperatures and solid angles.  These values are reported in Table \ref{excesslums}, along with the combined luminosity of the system, $L_{tot}=L_*+L_{acc}$, and the fraction of the total system luminosity emitted by the combined wall components, $L_{wall}=L_{cool}$, $L_{warm}$, or $L_{cool}+L_{warm}$ where applicable.  Since the star and accretion shock provide the total energy available to heat the physical structure in the circumstellar environment, including the wall, it is important that the wall have less luminosity than the total system luminosity; in all of our cases, this was true.  For all but the highest accreter, DR Tau, the wall has between 2 and 10\% of the total luminosity.  For DR Tau it is 17\%. We also confirmed that the luminosity of the hot component did not exceed $L_{acc}$ for any of these stars.  In the next section, we discuss the implications of these luminosities as well as the characteristic temperatures and solid angles within the context of their physical structure and composition.

\section{Discussion}
\label{discussion}

\subsection{Infrared Spectral Typing}
\label{irspt}
Based on our comparison of trends between equivalent width and spectral type in T Tauri, dwarf, and giant stars, we determined the degree to which individual absorption lines seen in our TTS were affected by surface gravity, finding that  in the majority of lines between 1 and 2.5 $\mu$m the T Tauri stars lie at an intermediate ($W_{\lambda}$, SpT) position between the dwarf and giants.  This is consistent with the results of \citet{lr98}, who found it necessary to interpolate between the dwarf and giant trends in order to fit well their TTS at $K$ band.  In general, for the deeper absorption features the dwarf trend has higher equivalent widths than the giants, so if one simply uses the dwarf trends to determine spectral types for TTS, the TTS will appear to have later spectral types in the IR than in the optical.  Such an effect has been noted by several authors, e.g. \citet{gullbring+98b}.

An additional factor contributing to a given star having an apparently later spectral type in the infrared is the effect of star spots, for which we see some evidence.  Many of the lines that peak at late M-types show TTS equivalent widths that are larger than predicted given their optical spectral type. In some cases, e.g. the 1.25 $\mu$m K I line, there are no differences between the dwarf and giant $W_{\lambda}$ vs SpT trends over our spectral type range, suggesting that the observed effect is not related to differences in surface gravity but instead to the presence of cooler spots on the stellar surface.  

\begin{figure}
\includegraphics[angle=0, scale=0.6]{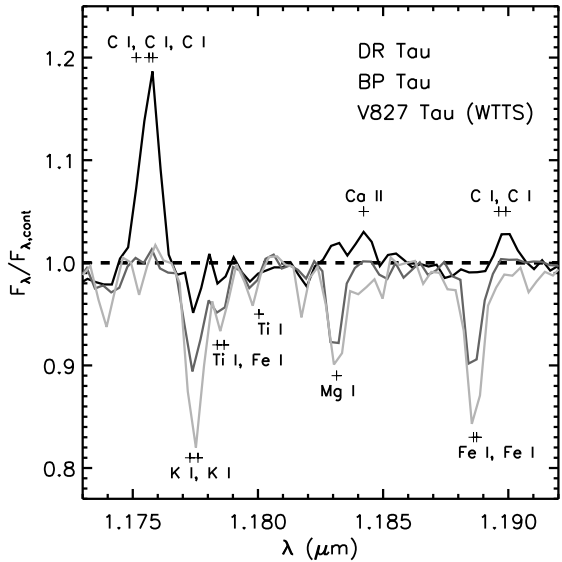}
\caption{Probable chromospheric emission around the 1.183 $\mu$m feature.  The T Tauri stars are V827 Tau (K7.5 WTTS, light grey), BP Tau (K7.5 CTTS, dark grey), and DR Tau (K7 CTTS, black).  Wavelengths corresponding to absorption and emission features are indicated by a `+' and labelled with their major absorber.   \label{chrom18}}
\end{figure}

\begin{figure}
\includegraphics[angle=0, scale=0.6]{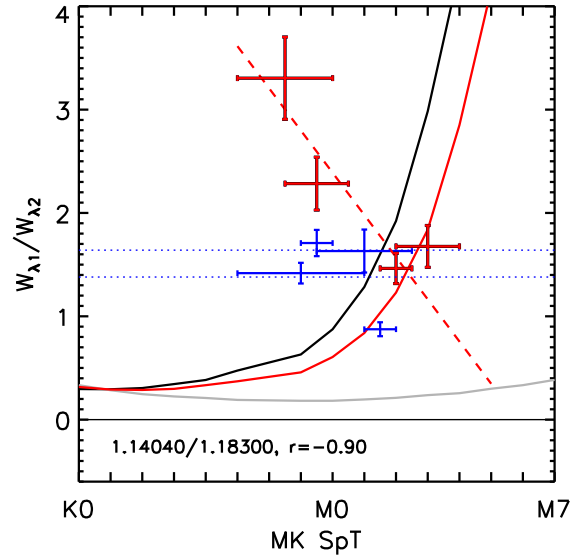}
\caption{Equivalent width ratio between 1.1404 (Na I) and 1.183 (Mg I) lines, following \citet{vacca11}.   Black curve is best-fitting ratio trend for the dwarf standards, grey curve is best-fitting curve for the giant standards, and red curved is the best-fitting curve for TTS (assuming that the TTS are supposed to lie on the dwarf trend for Mg I, barring chromospheric or other effects).  The red, dashed line represents the linear fit to the TTS trend based on where the WTTS actually lie in $W_{\lambda}$ vs. SpT diagram for the 1.183 $\mu$m feature.  Red crosses are the WTTS, assuming their spectral types as determined from optical data, while blue crosses are the CTTS with spectral types determined in this analysis.  Blue dashed lines are the approximate $W_{\lambda}$ ratio for TW Hya, as given by \citet{vacca11}, after dividing their Na I $W_{\lambda}$ in half, as we only measure the longer wavelength line in the doublet.  We note that \emph{all} of the K7 TTS would be assigned spectral types M1 to M3 using this diagnostic, based on their ratios of 1.4 to 1.65.  \label{rat18}}
\end{figure}

We also note that the effects of differential veiling or spectrally unresolved contributions from nearby chromospheric emission lines considerably complicate the analysis of lines shortwards of 1.2 $\mu$m.  For example, in the region surrounding the 1.183 $\mu$m Mg I line (Fig. \ref{chrom18}) the highest accreters show emission lines at wavelengths corresponding to C I and Ca II absorption lines in the solar spectrum \citep[See Table 6 of][]{rayner09}.  In DR Tau, there is even a suggestion of emission from the Mg I line itself, although our spectral resolution is too low to confirm this.  The result of this, as discussed in \S \ref{regveiling} is that almost all of the TTS have $W_{1.183}$ consistent with M2 to M3 dwarf spectral types, regardless of their optical spectral type.  Only three of the TTS, GO Tau, FN Tau, and LkCa3 have an equivalent width for this feature that is consistent with the dwarf trend.  This effect, combined with the likelihood that star spots contribute to the Na I 1.14040 line likely produce the result of \citet{vacca11}, who use a set of four late-M lines plus the Mg I 1.1833 line to reclassify TW Hydra from a K7 (optical) to M2 (infrared).  When we consider their preferred line ratio for spectral typing, Na I/Mg I (Fig. \ref{rat18}), we find that \emph{all} of the CTTS except for GO Tau appear to be $\sim$M2 by this measure, and the K7 WTTS appear to be even later.  Whatever the exact causes, this ratio is not advisable for determining spectral types.   We emphasize here how important it is to exercise caution when using dwarf standards to interpret TTS line equivalent widths for this reason.

In general, when veiling is taken into account through equivalent width ratios, the near-infrared is an ideal wavelength range over which to measure spectral types for K and M stars.  The stellar photospheres are at a maximum near $z$ and $J$ bands, and the continuum veiling due to the accretion shock and/or wall excess is at a minimum between $z$ and $H$ bands.  We note, however, that $H$ band may be superior to $J$ for the highest accreting stars, due both to the decrease in differential veiling from lines in the accretion shock and because the bump in $H$ band due to the lower surface gravity in TTS means that late TTS photospheres are almost as bright at $H$ as $J$.  Therefore, for more extinguished stars, one can achieve better S/N at $H$.

\subsection{Emission Size Scales}
\label{sizes}
Since we can place only tenuous constraints on the hot component properties due to the lack of spectral coverage shortwards of 0.8 $\mu$m, we simply note that for almost all of the accreters the hot component solid angles are on the order of 1 to 3\% of the stellar solid angle, while the luminosity of this component is always less than the accretion luminosity derived from Br$\gamma$ (Tables \ref{results} and \ref{excesslums}).  Therefore the hot component is not inconsistent with an accretion shock, particularly as shock emission is known to diverge from a single $\sim$8000 K blackbody spectrum, with a significant amount emission in the UV or X-ray \citep{cg98,ingleby+09}.  

In contrast, we can place firm lower limits on the emission area of the wall, by making the assumption that the wall emission is coming from a grey, optically thick dust.   In that case, and assuming the wall to be avertical surface, we can use the total system luminosity and $T_{wall}$ to determine a crude estimate of the dust destruction radius:

\begin{equation}
R_{wall}=\left( \frac{L_*+L_{acc}}{4\pi\sigma_b T_{wall}^4} \right)^{1/2}
\label{rwall}
\end{equation}

Our estimates of the radii of blackbody emission are given in units of the stellar radius and in AU in Table \ref{radheights} and are on the order of 0.1 AU or less.  We note that our radii estimates are slightly smaller than what has been found using interferometry for BP Tau, CI Tau, and DR Tau \citep{akeson+05b, eisner+07,akeson+05a}.  As seen in Table \ref{littab}, the inner radii for these disks is found to be $\sim$0.07 to 0.1 AU, assuming a ring model, while our blackbody radii for these disks are all 0.06 to 0.08 AU.  

\begin{figure}
\includegraphics[angle=0, scale=0.4]{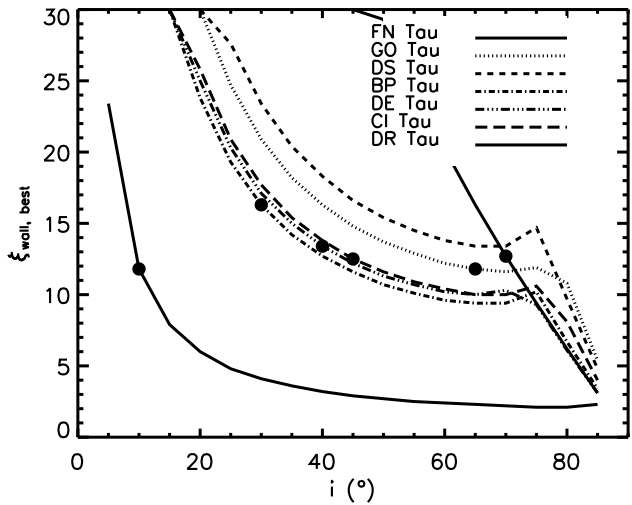}
\caption{Approximate best-fitting scale-factors, $\xi$, for the height, $z=\xi H$, in terms of the disk pressure scale height, $H$, as a function of inclination angle, $i$, for the six program stars whose excesses fits included a warm blackbody around $\sim$1600 to 1800 K.  Names of each star are given in plot.  Solid circles indicate the inclination attributed to the systems, where available.    \label{xivsi}}
\end{figure}

Our wall radii are calculated assuming grey dust, reasonable for millimeter sized grains.  If the grains are smaller, their opacity at shorter wavelengths increases, and they will reach their sublimation temperature at larger radii.   One way to test if the grey dust, or blackbody, assumption is reasonable is to compare the height of a wall at the blackbody radius given the solid angle found from the parametric fits, $\Omega_{wall}$, to the expected `surface' height of gas at the wall temperature and radius, $z_s=4H$.  The height, $z$, of a vertical wall with area $2\pi Rz$ is expressed as $z=\xi H$, where $\xi$ is a scaling factor and $H$ is the pressure scale height of the gas at that particular radius, given by  \citep{dalessio98}:

\begin{equation}
H=R^{3/2}\left( \frac{kT_c}{GM_*\mu(T_c,\rho_c)m_H} \right)^{1/2}
\label{scaleheight}
\end{equation}

\noindent where $T_c$ and $\rho_c$ are the temperature and density of the midplane.  In our case, $R=R_{wall}$, $T_c=T_{wall}$, and $\mu(T_c,\rho_c)=2.34$ under the assumption that the midplane is predominately molecular.  We calculate $H$ for each of the stars and list the results in Table \ref{radheights}.  Next, we estimate $\xi$, which is inclination dependent.  Assuming the wall is vertical, it can be approximated by a cylinder of radius $R_{wall}$ centered on the star, as discussed in Appendix B of \citet{dullemond+01}.  We use the prescription given there with a grid of angles between 0 and 90\degr to identify the value of $\xi$ for which the projected area best matches the solid angle of the cool blackbody as a function of $i$ (see Fig. \ref{xivsi}).  For the stars with known inclination angles, we overplot the best-fitting $\xi$ in Fig. \ref{xivsi} and record these values in Table \ref{radheights}.  

The best-fitting $\xi$ are between 10 and 20. These values are unrealistically high, i.e. much greater than that of the disk photosphere, typically $\sim$4. This suggests that although a composite of blackbody fits is able to reproduce the excess flux, the majority of the emission cannot be from truly blackbody grains.   If the grains are small enough not to have grey opacities, their radii would be larger than those given by these blackbody estimates (for the same value of the solid angle) and therefore the height of the wall required to match the solid angle would obviously decrease.  The requirement that the grains be large enough to have approximately blackbody shaped emission while small enough not to have grey emission should place relatively narrow constraints on the maximum grain size in the wall for future physical models. 

\subsection{Gas in the Inner Disk vs. a Curved Wall}
Our blackbody radii are less than a factor of two less than the radii estimated from ring interferometry models.  Several authors \citep{eisner+07,ajay+08,eisner+09} have suggested that these small disk radii indicate that the emission comes from optically thick gas inside the wall, based on their assumption of a particular disk temperature structure and SED/visibility fitting.  \citet{fischer+11} find an $H$ band excess in DR Tau that they fit with component at $\sim$2500 to 4000 K, temperatures which would be consistent with those predicted for optically thick gas.  While our present data cannot be compared with visibilities, in our simple, parametric analysis of the excess we found no evidence for a component at $\sim$2500 to 4000 K.  In all cases, the best-fitting models had a blackbody fitting the bulk of the $H$ and $K$ band excess with a temperature $<$2000 K.  This temperature is consistent with the most refractory grains.  

Given that the temperature of the disk at the dust sublimation radius depends on the opacity of the dust in the wall, it is not straightforward to say that a particular radial location in the disk is `too hot' for dust without exploring the grain properties.  Refractory dust (e.g. iron-, calcium-, or aluminum-rich silicates) have been found in solar-system meteorites.  There are suggestions of radial gradients of dust composition in these meteorites and other solar system bodies (i.e. more iron near Mercury's location).  A gradient in either grain size or composition could produce structure in the wall, with the more refractory grains could potentially exist closer to the star, creating curvature.  For at least one target with continuum emission from `inside' the dust-sublimation radius, follow-up observations by \citet{najita+09} suggest that the detected excess is unlikely to be be gaseous, but may instead be highly refractory dust.  Alternatively, curvature in the wall structure could be induced by taking into account how the dust sublimation temperature changes for different local pressures or how the density structure changes when dust settling is included, as do \citet{isellanatta05} and \citet{ajay+07}, respectively.  

This option is particularly interesting in light of the $\sim$800 to 1000 K blackbody component in FN Tau, DS Tau, CI Tau, and DR Tau, which required two blackbodies to fit the wall emission in our parametric analysis.  If one takes the temperatures of the warm and cool blackbodies as characteristic of the same physical structure, they may correspond to the inner and outer edges of the dust-sublimation front, where the warm component is in the high density, higher dust sublimation temperature midplane, and the cool component is in the lower density, lower dust sublimation temperature upper layers, \`{a} la \citet{isellanatta05}.  On the other hand, if one takes into consideration the effects of grain size, our data appear similar to the excess SEDs computed by \citet{ajay+07} for their dust-segregation model.

An additional complication for using SED fitting to suggest gaseous emission is the potential for a substantial contribution from scattered light from the disk around 1 to 2 $\mu$m, again depending on the grain properties. For late K and early M stars, this contribution would have a temperature of 3,500 to 4000 K.  Full disk models including radiative transfer, grain size distributions, and a variety of grain compositions are needed in order to determine how large a contribution this is.  Such models would also enable us to test the radial of the wall in a more physically realistic way.  We address this in the second paper in this series, McClure et al. 2013b (Paper II).

\section{Conclusions}

We have combined several techniques from the literature, e.g. veiling independent spectral-typing, determination of $A_V$ from veiling, and extracting excess emission from veiling, to determine self-consistently basic, model-independent properties relating to the star and NIR emission excess for a small sample of T Tauri stars in Taurus.  From this work, our main conclusions are:

\begin{itemize}
\item The later spectral types and colors often found for infrared classifications of T Tauri stars are due to differences between the photospheres of TTS and dwarf or giant standards, e.g. surface gravity.  Interpolation between the dwarf and giant trends to fit the TTS allows us to correct for the surface gravity of the stars to obtain spectral types consistent with those in the optical and more consistent veilings.
\item The 0.8 to 5 $\mu$m excess in T Tauri stars can be modeled successfully without appealing to emission from an inner, gaseous disk.  Instead, we find evidence for emission at temperatures {\it cooler} than the dust sublimation temperature. 
\item The additional $\sim$ 800 K blackbody required to fit the 3 to 5$\mu$m excesses of three of our program stars may be evidence for curvature in their sublimation walls or a contribution near 5 $\mu$m from the disk surface beyond the wall.
\item The solid angles of the NIR excess are large, and the wall heights required to match them (using a blackbody approximation to calculate the wall radius) are an order of magnitude too large.  Therefore the wall must be populated with grains that are small enough not to have grey opacities in the NIR, e.g. less than tens of microns.
\item We explore the latter two conclusions further with a physical treatment of the wall and disk in Paper II.
\end{itemize}

Aside from these conclusions, the analysis here demonstrates the importance of obtaining simultaneous, moderate resolution spectra over a wide span of wavelengths to self-consistently determine the properties of both young stars and their excesses.  It also underscores the importance of obtaining high quality spectra of single, weakly-accreting T Tauri stars of known optical spectral types at infrared wavelengths to use as templates for the classical T Tauri stars in one's sample.  While in general WTTS are similar enough to dwarf standards of the same optical spectral type, at infrared wavelengths, there are enough differences in surface gravity and chromospheric activity that neither dwarfs nor giants are as good as photospheric templates as a WTTS.  As we attempt to unveil the excess properties of the inner regions of disks around CTTS, it is essential to properly constrain the underlying photosphere.

For now, these results are suggestive but preliminary.  It is necessary to study a much larger sample of stars using physical models of the accretion shock and wall to understand the details of the infrared excesses in these systems.  Doing so is worthwhile, however, to broaden our understanding of the energetics and dust evolution of the innermost regions of circumstellar disks and how they impact the terrestrial planet-forming region.

\clearpage

\appendix

\section{A.  Trends in $W_{\lambda}$ vs SpT}
\begin{figure}
\includegraphics[angle=0, scale=0.7]{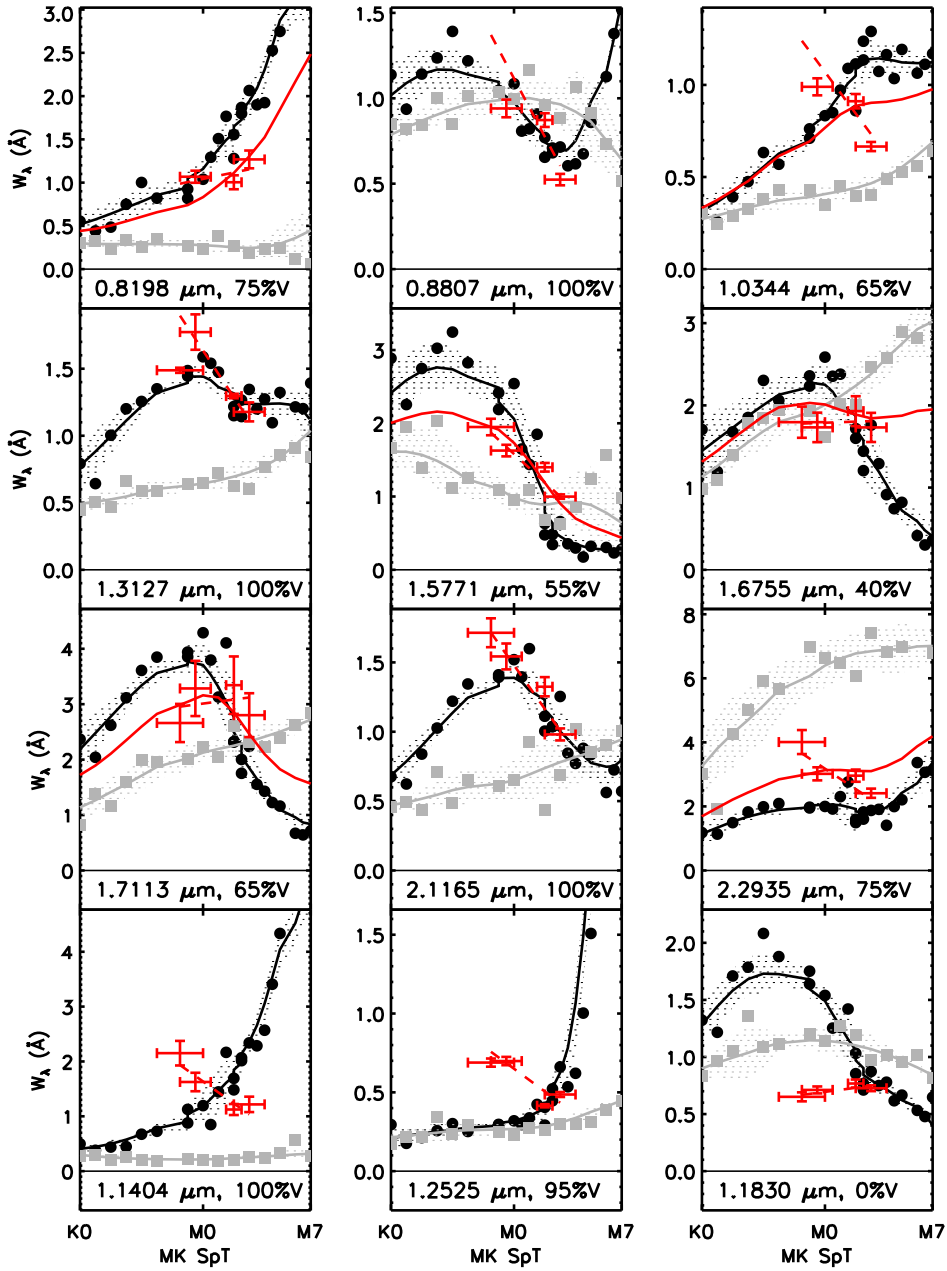}
\caption{{\it Top three rows:} Trends in $W_{\lambda}$ of surface gravity sensitive lines as a function of spectral type for our WTTS and weakly veiled CTTS (positions indicated by red error bars) and the IRTF spectral library dwarf (black filled circles) and giant (grey filled squares) standards.  Fits to dwarf and giant trends are given by solid lines, with uncertainties indicated by the hatched fill in the appropriate color.  The best-fitting interpolation between the dwarf and giant trends for the TTS is also plotted (solid red line).  Fraction of the dwarf trend contributed to the interpolation is given in the bottom right corner (see text). {\it Bottom row:} Trends in $W_{\lambda}$ of potentially star-spot (1$^{st}$ and 2$^{nd}$ panels) or chromospheric (last panel) lines as a function of spectral type for our WTTS and weakly veiled CTTS.}
\label{gravtrendsa}
\end{figure}

For each line in our, we compared the equivalent widths of the TTS with those of the dwarf and giant standard stars in the IRTF spectral type library \citep{rayner09}.  The library contains at least one standard star of luminosity classes V and III per spectral type between F0 and M9.  For each line, we produced a plot of the equivalent width of that line as a function of spectral type for both the dwarf and giant standards in the IRTF library.  To fit each trend, we used a non-parametric locally weighted scatterplot smoothing (LOWESS) algorithm \citep{cleveland_devlin88} to smooth the data in bins of four spectral subtypes and computed an uncertainty in our fit.  Several of the lines for which the equivalent width depended strongly on the spectral type are shown in Fig. \ref{gravtrendsa}.

In the panel for each line, we also overplot the location of our two WTTS and two weakly veiled CTTS, including error bars on both their $W_{\lambda}$ and on their spectral type, assuming the spectral types determined from optical data using the Hernandez et al. SPTCLASS package.  There are three types of trends.  In the first, the shape of the WTTS as a function of spectral type mirrors that of either the giants or dwarfs.  For these lines, the WTTS typically either lie directly on the dwarf curve or at values between those of the dwarfs and giants, consistent with the intermediate surface gravity of TTS.  These lines are typically metal lines with peak equivalent widths in the mid-G to K range.  For these lines we interpolate between the dwarf and giant trend as $x\times W_{\lambda,dwarf} + (1-x)\times W_{\lambda,giant}=W_{\lambda,TTS}$.  Examples of these TTS trends are plotted in the first three rows of Fig. \ref{gravtrendsa} and we report $100\times x$ in the lower right-hand corner of each panel of that figure.

However, for the second type of trend, seen in metal lines that peak at mid-M, e.g. 1.1404 (Na I) and 1.25250 (K I), our K7 WTTS and K6.5 weakly-veiled CTTS have equivalent widths consistent with M2 to M5 spectral types.  In particular, these lines have little or no surface gravity dependence over our spectral type range, and our other WTTS and weakly-veiled CTTS {\it do} lie on the dwarf curve at the location corresponding to their optical spectral type (Fig. \ref{gravtrendsa}, bottom-left two panels).  This situation is suggestive of star spots, which have spectral signatures consistent with cooler effective temperatures.  

The final trend is when all of the TTS lie below both the dwarf and giant trends, despite strong detections (e.g. Fig. \ref{gravtrendsa}, bottom-right panel).  In this case, there are emission lines near that location that appear in our highest accreters.  These lines corresponds to transitions seen in the solar spectrum.  See \S \ref{irspt} for more details.

\label{app2}

\section{B.  Spectral type determination from $W_{\lambda}$ ratios}

\begin{figure}
\includegraphics[angle=0, scale=0.7]{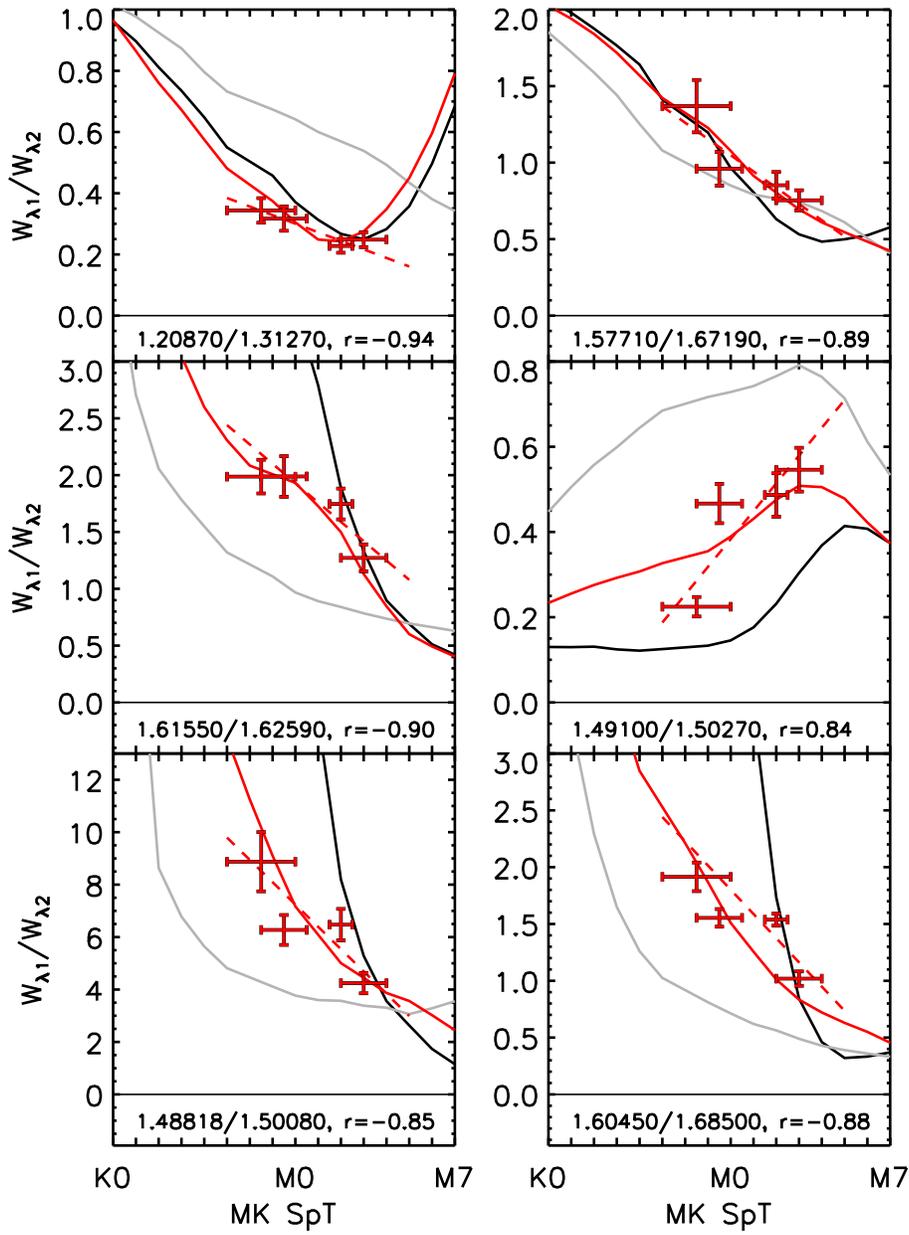}
\caption{Ratios of equivalent widths that we use for spectral typing.  Colors and symbols have the same meaning as in Fig. \ref{gravtrendsa}.  Lines over which the ratio is taken, as well as the Pearson correlation coefficient between the ratio and the SpT for the TTS, are given at the bottom of each panel.  \label{sptrat}}
\end{figure}

For spectral type determination, we took ratios between all pairs of lines within 0.1 $\mu$m of each other, computed the WTTS trend of the ratio, and then compared how well the WTTS ratio for a particular line pair correlated linearly with SpT over the K5 to M5 range. We then took the ratios with a Pearson correlation coefficient $r > 0.8$, i.e. strong correlations, and we then inspected them by eye.  We selected only those ratios for which both the dwarf and giant ratio trends showed little scatter and in which the WTTS trend shared the same shape as one or both of the standard trends.  In the end, we identified six independent pairs of lines that satisfy these criteria, which can be see in Fig. \ref{sptrat}.

We note, however, that even under these circumstances, some of the program stars do not fall on the WTTS trendline.  In particular, DR Tau, our strongest accreter, has emission lines near many of our feature pairs, so we are either unable to obtain good fits for some features or for others we may have chromospheric emission in part of the feature.  The lines and targets for which this occurred are noted in Table \ref{veilings}.

\label{app3}

\clearpage
\acknowledgments
This work is based on observations made with the NASA Infrared Telescope Facility.  This material is based upon work supported by the National Science Foundation Graduate Student Research Fellowship under Grant No. DGE 0718128.  N. C. acknowledges support from NASA Orgins grant NNX08AH94G.  K. L. was supported by grant AST-0544588 from the National Science Foundation.  The Center for Exoplanets and Habitable Worlds is supported by the Pennsylvania State University, the Eberly College of Science, and the Pennsylvania Space Grant Consortium.  This publication made use of NASA's Astrophysics Data System Abstract Service as well as the SIMBAD database and Vizier catalog service, operated by the Centre de Données astronomiques de Strasbourg.

\begin{deluxetable}{ccccc}
\tabletypesize{\small}   %\small (11pt), \footnotesize (10 pt), \scriptsize (8pt)
\tablewidth{0pt}
\tablecaption{Target List}
\tablehead{\colhead{Name} & \colhead{Obs. Type} & \colhead{K$_{S}$ $^a$}  & \colhead{K$_{MKO}$ $^b$} \\
 &  & \colhead{(mag)} & \colhead{(mag)}}
\startdata
{\it Standards} \\
FS 117 & photometric  & 10.05 $\pm$ 0.018 \\
HD 27761 & telluric  & 7.27 $\pm$ 0.020 \\
\hline \\
{\it Program stars} \\
V827 Tau$^c$ & WTTS &  8.230 $\pm$ 0.02 & 8.22 \\
LkCa3$^c$ & WTTS &  7.423 $\pm$ 0.02 &  7.41 \\
V836 Tau & CTTS & 8.595 $\pm$ 0.02 & 8.12 \\
FN Tau & CTTS &  8.189 $\pm$ 0.02  & 8.14 \\
GO Tau & CTTS &  9.332 $\pm$ 0.02 & 9.43 \\
DS Tau & CTTS &  8.036 $\pm$ 0.03 & 8.10 \\
BP Tau & CTTS & 7.736 $\pm$ 0.02 & 7.42 \\
DE Tau & CTTS & 7.799 $\pm$ 0.02 & 7.57 \\
CI Tau & CTTS &  7.793 $\pm$ 0.02 & 7.79 \\
DR Tau & CTTS & 6.874 $\pm$ 0.02 & 6.74
\enddata
\tablecomments{ \\
$^a$ $K_S$ magnitudes are taken from the 2MASS survey \citep{cutri+03}. \\
$^b$ $K_{MKO}$ measured in this work. Uncertainty on these magnitudes is $\sim$0.01 mag. \\
$^c$ Both WTTS are binaries.  At $K$, LkCa 3 (AB) has a separation of 0\farcs48 and a magnitude difference, $\delta$m, of 0.22 \citep{wg01}.  
At $H$, V827 Tau (AB) has a separation of 0\farcs09 and $\delta$m=0.58 magnitudes \citep{kraus+11}.
}
\label{sampletab}
\end{deluxetable}

\begin{deluxetable}{cccccc}
\tabletypesize{\small}   %\small (11pt), \footnotesize (10 pt), \scriptsize (8pt)
\tablewidth{0pt}
\tablecaption{System parameters from the literature}
\tablehead{\colhead{Name} & \colhead{SpT$^a$} & \colhead{$A_V$$^a$} & \colhead{$L_{*}$$^b$} & \colhead{$L_{acc}$$^b$} & \colhead{$R_{wall}$$^c$} \\
  &  & \colhead{(mag)} & \colhead{(\Lsun)}  & \colhead{(\Lsun)} & \colhead{(AU)}}
\startdata
V827 Tau  & K7   & 0.28  & 0.79  & ...  & ...   \\
LkCa3       & M1   & 0.42  & 1.7    & ...  & ...   \\
V836 Tau  & K7   & 0.59  & 0.51  & ...  &  ...  \\
FN Tau      & M5  & 1.35  & 0.7     & ...  & ...   \\
GO Tau     & M0   & 1.18  & 0.37  & 0.10  & ...   \\
DS Tau     & K5    & 0.31  & 0.58  & 0.21  & ...   \\
BP Tau     & K7    & 0.49  & 0.93  & 0.18  & 0.083$^{+0.029}_{-0.041}$, 0.123$^{+0.041}_{-0.055}$   \\
DE Tau     & M2   & 0.59  & 0.87  & 0.07  &  ...  \\
CI Tau       & K7   & 1.77  & 0.81  & 0.43  &  0.097 $\pm$ 0.008  \\
DR Tau     & K7   & ...       & 0.87  & 1.01  & 0.070 $\pm$ 0.026, 0.11 $\pm$ 0.041
\enddata
\tablecomments{ \\
$^a$ SpT and $A_V$ are from \citet{kh95}. \\
$^b$ Luminosities are from \citet{kh95}, \citet{hartmann+98} and \citet{muzerolle03}.  For V827 Tau, LkCa 3, V836 Tau, and FN Tau, the luminosities listed are bolometric. \\
$^c$  Radii are from interferometric modeling by \citet{akeson+05a}, \citet{akeson+05b}, and \citet{eisner+07}.  The first value in each entry is the radius of a ring model, while the second value (if present) is for a uniform disk model.  The model for CI Tau was assumed to be face-on.
}
\label{littab}
\end{deluxetable}

\begin{deluxetable}{llllllll}
\setlength{\tabcolsep}{0.04in}
\tabletypesize{\small}   %\small (11pt), \footnotesize (10 pt), \scriptsize (8pt)
\tablewidth{0pt}
\tablecaption{Veilings}
\tablehead{\colhead{$\lambda$ ($\mu$m)} & \colhead{Species} & \colhead{GO Tau} & \colhead{DS Tau} & \colhead{BP Tau}  & \colhead{DE Tau}  & \colhead{CI Tau}  & \colhead{DR Tau}}
\startdata
%						GO		DS			BP		DE		CI			DR
0.81980  	&	 Na I 			& 0	  	& $...^d$  	&	0.5	&	0.3	&	0.1		&	2.7	\\
0.88070  	&	 Na I 			& 0  		& $...^d$  	&	$^c$	&	0.3	&	$^c$		&	$^c$	\\
0.97880	&	Ti I 			& 0		& 0.4  	&	0.4	&	0.3	&	0.5		&	2.4	\\
0.99150	&	TiO			& 0		& 0.3		&	0.4	&	0.3	&	$...^c$	&	$...^c$\\
1.03440	&	Ca I 			& 0		& 0.3		&	0.4	&	0.4	&	0.6		&	2.8	\\
1.20870	&	Mg I/Si I		& 0		& 0.4 	&	0.4	&	0.4	&	0.6		&	2.6	\\
1.31270	&	Al I 			& 0		& 0.4		&	0.5	&	0.4	&	1.0		&	2.5	\\
%1.31544  	&	Al I 			& 0 		& 0.4		&	0.5	&	0.4	&	0.8		&	2.5	\\
1.50270  	&	Mg I 			& 0		& 0.3		&	0.4	&	0.4	&	$...^a$	&	$...^c$	\\
1.57721  	&	Mg I / Fe I 		& 0		& 0.5		&	0.4	&	0.4	&	0.8		&	2.7	\\
%1.61550	&	Ca I 			& 0		& 0.2		&	0.4	&	0.4	&	0.7		&	3	\\
%1.62590	&	Si I/ Fe I 		& 0		& 0.6		&	0.4	&	0.4	&	1.2		&	2.3	\\
1.67550  	&	Al I 			& 0		& 0.2		&	0.5	&	0.5	&	0.7		&	2.4	\\
1.71130  	&	Mg I 			& 0		& 0.3		&	0.4	&	0.4	&	0.8		&	3.4	\\
1.99340  	&	Fe I/Si I/Ca I 	& 0		& 0.4		&	0.6	&	0.7	&	  $...^a$  	&	3.2	\\
2.11650	&	Al I 			& 0.1		& 1.1		&	0.8	&	0.9	&	1.7		&	6.7	\\
%2.26270  	&	Ca I 			& 0.3		& 0.7		&	0.8	&	0.7	&	1.6		&	6.9	\\
2.29350	&	CO 			& 0.9		& 0.8		&	0.8	&	1.1	&	1.4		&	$...^b$	

\enddata

\tablecomments{The formal uncertainties on the non-zero measurements were typically 0.1, except for CI Tau (typically 0.2-0.4) and DR Tau (typically 0.5-1). \\
$^a$ Features are in regions of poor telluric correction. \\
$^b$ Features are below the continuum criterion of 2\%. \\
$^c$ Features are in the vicinity of multiple emission lines, preventing a good continuum fit. \\
$^d$ This line had poor signal-to-noise for this target.
}
\label{veilings}
\end{deluxetable}

\begin{deluxetable}{llllllll}
\tabletypesize{\small}   %\small (11pt), \footnotesize (10 pt), \scriptsize (8pt)
\tablewidth{0pt}
\tablecaption{Stellar parameters}
\tablehead{\colhead{Name} & \colhead{SpT} & \colhead{SpT} & \colhead{$A_V$$^a$} & \colhead{$T_{eff}$} & \colhead{$L_{*}$} & \colhead{$R_{*}$} & \colhead{$M_{*}$$^b$} \\
  & (optical) & (infrared)  & (mag) &  (K) & (\Lsun) & (\Rsun) & (\Msun) 
}
\startdata
LkCa 3       & M2.0 $\pm$ 0.5 &  ...                     &  0.5$\pm$0.1  &  3580  &  2.0$_{-0.1}^{+0.1}$  &  ...$^c$  &  ...$^c$   \\
V836 Tau  & K6.5 $\pm$ 1.5 &  ...                      &  1.4$\pm$0.5  &  4060  &  1.3$_{-0.5}^{+0.7}$  &  2.28  &  0.74   \\
FN Tau       & M3.0 $\pm$ 1.0 &   ...                    &  1.0$\pm$0.4  &  3470  &  0.8$_{-0.2}^{+0.4}$  &  2.46  &  0.35   \\
GO Tau      & M1.0 $\pm$ 0.5 &  M0$\pm$1.5  &  1.0$\pm$0.6  &  3850  &  0.3$_{-0.1}^{+0.2}$  &  1.13  &  0.59   \\
DS Tau      & K6.5 $\pm$ 1.0 &  K7 $\pm$1.5  &  1.4$\pm$0.5  &  4060  &  0.7$_{-0.3}^{+0.4}$  &  1.69  &  0.77   \\
BP Tau       & K6.5 $\pm$ 1.0 &  M0$\pm$1.5  &  0.6$\pm$0.2  &  3850  &  1.0$_{-0.2}^{+0.2}$  &  2.28  &  0.56   \\
DE Tau       & M1.5 $\pm$ 0.5 & M1$\pm$1.0  &  0.9$\pm$0.5  &  3720  &  0.9$_{-0.3}^{+0.5}$  &  2.29  &  0.47   \\
CI Tau        &  K5.5 $\pm$ 1.0 & K7 $\pm$1.0  &  1.3$\pm$0.8  &  4060  &  0.6$_{-0.3}^{+0.7}$  &  1.57  &  0.78   \\
DR Tau       & K3.0 $\pm$ 2.0 &  M0$\pm$1.5  &  2.0$\pm$1.3  &  3850  &  0.6$_{-0.4}^{+1.5}$  &  1.79  &  0.57   
\enddata
\tablecomments{
\\
$^a$ Uncertainties on $A_V$ are 3$\sigma$.  \\
$^b$ As given by the \citet{siess2000} PMS evolutionary tracks, using our derived $T_{eff}$ and $L_{*}$ as input. \\
$^c$ Because both WTTS are binary, we do not report stellar masses for them. }
\label{starparams}
\end{deluxetable}

\begin{deluxetable}{llllllll}
\tabletypesize{\small}   %\small (11pt), \footnotesize (10 pt), \scriptsize (8pt)
\tablewidth{0pt}
\tablecaption{Excess parameters}
\tablehead{\colhead{Name} & \colhead{$T_{cool}$}  & \colhead{$\Omega_{cool}$}  & \colhead{$T_{warm}$}  & \colhead{$\Omega_{warm}$}  & \colhead{$T_{hot}$} & \colhead{$\Omega_{hot}$} & \colhead{$\chi^2$/d.o.f.} \\
  & \colhead{(K)} & \colhead{($\Omega_{*}$)} & \colhead{(K)} & \colhead{($\Omega_{*}$)} & \colhead{(K)} & \colhead{($\Omega_{*}$)}
}
\startdata
V836 Tau$^a$  &  ...  &  ...  & 1200  & 10.9  &    8000  &  0.005 & 4.42  \\
FN Tau  &  900  &  33.8  &  1700 & 1.4 &  ...  &  ...  &  10.97 \\
GO Tau  &  ...  &  ...  &  1650  &  4.6  &  8000  &  0.001  & 3.3 \\
DS Tau  &  900  &  20.1  &  1700  &  6.2  &  8000  &  0.026 &  3.82  \\
BP Tau  &  ...  &  ...  &  1700  &  7.9  &  6000  &  0.051 & 9.81   \\
DE Tau  &  ...  &  ...  &  1600  &  10.5  &  8000  &  0.007 &  18.19  \\
CI Tau  &  1000  &  30.0  &  1800  &  12.3  &  7000  &  0.02 & 2.49   \\
DR Tau  &  800  &  277.5 &  1700  &  52.1  &  8000  &  0.165 & 10.74 
\enddata
\tablecomments{
\\
$^a$ V836 Tau was not observed with LXD, so the temperature is an upper limit and the solid angle is not well constrained. }
\label{results}
\end{deluxetable}

\begin{deluxetable}{llllllllll}
\tabletypesize{\small}   %\small (11pt), \footnotesize (10 pt), \scriptsize (8pt)
\tablewidth{0pt}
\tablecaption{Luminosities}
\tablehead{\colhead{Name} & \colhead{$L_{Br\gamma}$} & \colhead{$L_{acc}$} & \colhead{\Mdot} & \colhead{$L_{cool}$}  & \colhead{$L_{warm}$}   &  \colhead{$L_{hot}$}   & \colhead{$L_{tot}^b$}  & \colhead{$L_{wall}/L_{tot}$} & \colhead{$L_{hot}/L_{acc}$} \\
   & \colhead{(\Lsun)} & \colhead{(\Lsun)} & \colhead{(\Msun/yr)} & \colhead{(\Lsun)}  & \colhead{(\Lsun)}  & \colhead{(\Lsun)}  & \colhead{(\Lsun)}  & \colhead{(\Lsun)}  & \colhead{(\Lsun)}
}
\startdata
V836 Tau$^a$  &  ...  &  ... & ... & ...  &  0.03  &  0.02  &  1.97  &  0.02 & ...  \\
FN Tau  &  ...  &  ...  & ...  &  0.03  &  0.02 &  ...  &  0.79  &  0.06 & ...    \\
GO Tau  &  ...  &  ...  & ... &  ...  &  0.01  &  ...  &  0.25  &  0.04 & ...  \\
DS Tau  &  4.3e-5  &  0.08  & 5.6e-9 &  0.01  &  0.03  &  0.07  &  0.78  &  0.05 & 0.88 \\
BP Tau  & 5.8e-5  &  0.12  & 1.6e-8  &  ...  &  0.08  &  0.09  &  1.15  &  0.07 & 0.75 \\
DE Tau  &  6.8e-5  &  0.15  & 2.3e-8 &  ...  &  0.08  &  0.03  &  1.05  &  0.08 & 0.20   \\
CI Tau  &  1.8e-4  &  0.51  & 3.3e-8 &  0.02  &  0.07  &  0.03  &  1.12  &  0.08 & 0.06  \\
DR Tau  &  4.7e-4  &  1.71  & 1.7e-7 &  0.08  &  0.31  &  0.49  &  2.35  &  0.17 & 0.29  
\enddata
\tablecomments{
\\
$^a$ V836 Tau was not observed with LXD, so the temperature is an upper limit and the solid angle is poorly constrained. \\
$^b$ $L_{tot}=L_*+L_{acc}$ \\
$^c$ $L_{wall}=L_{cool}+L_{warm}$ or $L_{cool}$ or $L_{warm}$, depending on which combination of those two components are present. }
\label{excesslums}
\end{deluxetable}

\begin{deluxetable}{lllll}
\tabletypesize{\small}   %\small (11pt), \footnotesize (10 pt), \scriptsize (8pt)
\tablewidth{0pt}
\tablecaption{Wall radii and heights}
\tablehead{\colhead{Name} & \colhead{$R_{warm}$}  & \colhead{$R_{warm}$}  & \colhead{$H/R_{warm}$ $^a$} &\colhead{$\xi$}$^b$ \\
  & \colhead{($R_*$)} & \colhead{(AU)}
}
\startdata
V836 Tau & 11.5 & 0.121 & 0.028 & ... \\
FN Tau	& 4.2 & 0.048 & 0.031 & 11.8 \\
GO Tau  &  5.4 & 0.029 & 0.016 & 11.8 \\
DS Tau  &  5.4 & 0.046 & 0.018 & ... \\
BP Tau  &  5.4 & 0.057 & 0.026 & 16.3 \\
DE Tau  &  5.8 & 0.062 & 0.029 & 13.4 \\
CI Tau  &  6.2 & 0.050 & 0.022  & 12.5 \\
DR Tau  &  9.9 & 0.082 & 0.029 &  12.7
\enddata
\tablecomments{
\\
$^a$ $H$ is the gas pressure scale height at a given radius, as defined in Equation (\ref{scaleheight}). \\
$^b$ $\xi=z/H$ is calculated only for the stars with well constrained disk inclinations and warm temperature component fits.}
\label{radheights}
\end{deluxetable}


\begin{thebibliography}{}
\bibitem[Akeson et al.(2005a)]{akeson+05a} Akeson, R.~L., Walker, C.~H., Wood, K., et al.\ 2005, \apj, 622, 440 
\bibitem[Akeson et al.(2005b)]{akeson+05b} Akeson, R.~L., Boden, A.~F., Monnier, J.~D., et al.\ 2005, \apj, 635, 1173 
\bibitem[Andrews \& Williams(2005)]{aw05} Andrews, S.~M., \& Williams, J.~P.\ 2005, \apj, 631, 1134 
\bibitem[Andrews \& Williams(2007)]{aw07} Andrews, S.~M., \& Williams, J.~P.\ 2007, \apj, 659, 705 
\bibitem[Basri \& Batalha(1990)]{bb90} Basri, G., \& Batalha, C.\ 1990, \apj, 363, 654 
\bibitem[Batalha \& Basri(1993)]{bb93} Batalha, C.~C., \& Basri, G.\ 1993, \apj, 412, 363 
\bibitem[Calvet \& Gullbring(1998)]{cg98} Calvet, N., \& Gullbring, E.\ 1998, \apj, 509, 802 
\bibitem[Calvet \& D'Alessio(2011)]{cd11} Calvet, N., \& D'Alessio, P.\ 2011, Physical Processes in Circumstellar Disks around Young Stars, 14 
\bibitem[Cleveland \& Devlin(1988)]{cleveland_devlin88} Cleveland, W. S. \& Devlin, S. J., 1988, Journal of the American Statistical Association, 83, 403, 596
\bibitem[Cushing et al.(2005)]{cus05} Cushing, M. C., Rayner, J. T., \& Vacca, W. D. 2005, \apj, 623, 1115
\bibitem[Cushing et al.(2004)]{cus04} Cushing, M. C., Vacca, W. D., \& Rayner, J. T. 2004, \pasp, 116, 362
\bibitem[Cutri et al.(2003)]{cutri+03} Cutri, R.~M., Skrutskie, M.~F., van Dyk, S., et al.\ 2003, VizieR Online Data Catalog, 2246, 0 
\bibitem[D'Alessio et al.(1998)]{dalessio98} D'Alessio, P., Canto,  J., Calvet, N., \& Lizano, S.\ 1998, \apj, 500, 411 
\bibitem[D'Alessio et al.(2004)]{dalessio+04} D'Alessio, P., Calvet, N., Hartmann, L., Muzerolle, J., \& Sitko, M.\ 2004, Star Formation at High Angular Resolution, 221, 403 
\bibitem[D'Alessio et al.(2005)]{dalessio+05} D'Alessio, P., Hartmann, L., Calvet, N., et al.\ 2005, \apj, 621, 461 
\bibitem[D'Alessio et al.(2006)]{dalessio+06} D'Alessio, P., Calvet, N., Hartmann, L., Franco-Hern{\'a}ndez, R., 
\& Serv{\'{\i}}n, H.\ 2006, \apj, 638, 314 
\bibitem[Dorschner et al.(1995)]{dorschner+95} Dorschner, J., Begemann, B., Henning, T., Jaeger, C., \& Mutschke, H.\ 1995, \aap, 300, 503 
\bibitem[Draine \& Lee(1984)]{dl84} Draine, B.~T., \& Lee, H.~M.\ 1984, \apj, 285, 89 
\bibitem[Dullemond et al.(2001)]{dullemond+01} Dullemond, C.~P., Dominik, C., \& Natta, A.\ 2001, \apj, 560, 957 
\bibitem[Dullemond \& Dominik(2004)]{dd04} Dullemond, C.~P., \& Dominik, C.\ 2004, \aap, 417, 159 
\bibitem[Edwards et al.(2006)]{edwards06} Edwards, S., Fischer, W., Hillenbrand, L., \& Kwan, J.\ 2006, \apj, 646, 319 
\bibitem[Eisner et al.(2007)]{eisner+07} Eisner, J.~A., Chiang, E.~I., Lane, B.~F., \& Akeson, R.~L.\ 2007, \apj, 657, 347 
\bibitem[Eisner et al.(2009)]{eisner+09} Eisner, J.~A., Graham, J.~R., Akeson, R.~L., \& Najita, J.\ 2009, \apj, 692, 309 
\bibitem[Espaillat et al.(2010)]{espaillat+10} Espaillat, C., D'Alessio, P., Hern{\'a}ndez, J., et al.\ 2010, \apj, 717, 441 
\bibitem[Fabricant et al.(1998)]{fabricant+98} Fabricant, D., Cheimets, P., Caldwell, N., \& Geary, J.\ 1998, \pasp, 110, 79 
\bibitem[Finkenzeller \& Basri(1987)]{fb87} Finkenzeller, U., \& Basri, G.\ 1987, \apj, 318, 823 
\bibitem[Fischer et al.(2011)]{fischer+11} Fischer, W., Edwards, S., Hillenbrand, L., \& Kwan, J.\ 2011, \apj, 730, 73 
\bibitem[Furlan et al.(2006)]{furlan+06} Furlan, E., Hartmann, L., Calvet, N., et al.\ 2006, \apjs, 165, 568 
\bibitem[Garrison(1994)]{garrison94} Garrison, R.~F.\ 1994, The MK Process at 50 Years:  A Powerful Tool for Astrophysical Insight, 60, 3
\bibitem[Ghez et al.(1993)]{ghez+93} Ghez, A.~M., Neugebauer, G., \& Matthews, K.\ 1993, \aj, 106, 2005 
\bibitem[Gray \& Corbally (2009)]{gc09} Gray, R. O., \& Corbally, C. J.\ 2009, Stellar Spectral Classification (Princeton, NJ: Princeton University Press)
\bibitem[Gullbring et al.(1998a)]{gullbring+98a} Gullbring, E., Hartmann, L., Briceno, C., \& Calvet, N.\ 1998, \apj, 492, 323 
\bibitem[Gullbring et al.(1998b)]{gullbring+98b} Gullbring, E., Hartmann, L., Briceno, C., Calvet, N., \& Muzerolle, J.\ 1998, Cool Stars, Stellar Systems, and the Sun, 154, 1709 
\bibitem[Guilloteau et al.(2011)]{guilloteau+11} Guilloteau, S., Dutrey, A., Pi{\'e}tu, V., \& Boehler, Y.\ 2011, \aap, 529, A105 
\bibitem[Harder \& Schubert(2001)]{hs01} Harder, H., \& Schubert, G.\ 2001, Icarus, 151, 118 
\bibitem[Hartigan et al.(1989)]{hartigan+89} Hartigan, P., Hartmann, L., Kenyon, S., Hewett, R., \& Stauffer, J.\ 1989, \apjs, 70, 899 
\bibitem[Hartigan et al.(1995)]{heg95} Hartigan, P., Edwards, S., \& Ghandour, L.\ 1995, \apj, 452, 736 
\bibitem[Hartmann et al.(1998)]{hartmann+98} Hartmann, L., Calvet, N., Gullbring, E., \& D'Alessio, P.\ 1998, \apj, 495, 385 
\bibitem[Hawarden et al.(2001)]{hawarden+01} Hawarden, T.~G., Leggett, S.~K., Letawsky, M.~B., Ballantyne, D.~R., \& Casali, M.~M.\ 2001, \mnras, 325, 563 
\bibitem[Hemley et al.(1994)]{hemley94} Hemley, R.~J., Prewitt, C.~T., \& Kingma, K.~J.\ (1994) High-Pressure Behavior of Silica. In {\it Silica: 
Physical Behavior, Geochemistry and Materials Applications}, Reviews in Mineralogy (eds. P.~J.\ Heaney, C.~T.\ Prewitt, \& G.~V. Gibbs). 
Mineralogical Society of America, Washington, D.~C., vol. 29, chap. 2, pp 41-81
\bibitem[Herbst et al.(1994)]{herbst+94} Herbst, W., Herbst, D.~K., Grossman, E.~J., \& Weinstein, D.\ 1994, \aj, 108, 1906 
\bibitem[Hern{\'a}ndez et al.(2004)]{hernandez+04} Hern{\'a}ndez, J., Calvet, N., Brice{\~n}o, C., Hartmann, L., \& Berlind, P.\ 2004, \aj, 127, 1682 
\bibitem[Herczeg et al.(2004)]{herczeg+04} Herczeg, G.~J., Wood, B.~E., Linsky, J.~L., Valenti, J.~A., \& Johns-Krull, C.~M.\ 2004, \apj, 607, 369 
\bibitem[Ingleby et al.(2009)]{ingleby+09} Ingleby, L., Calvet, N., Bergin, E., et al.\ 2009, \apjl, 703, L137 
\bibitem[Isella \& Natta(2005)]{isellanatta05} Isella, A., \& Natta, A.\ 2005, \aap, 438, 899 
\bibitem[Ita et al.(2010)]{ita+10} Ita, Y., Matsuura, M., Ishihara, D., et al.\ 2010, \aap, 514, A2 
\bibitem[Jaeger et al.(1994)]{jaeger+94} Jaeger, C., Mutschke, H., Begemann, B., Dorschner, J., \& Henning, T.\ 1994, \aap, 292, 641 
\bibitem[J{\"a}ger et al.(2003)]{jager+03} J{\"a}ger, C., Dorschner, J., Mutschke, H., Posch, T., \& Henning, T.\ 2003, \aap, 408, 193 
\bibitem[Johns-Krull(2007)]{jk07} Johns-Krull, C.~M.\ 2007, \apj, 664, 975 
\bibitem[Kenyon et al.(1994)]{kenyon+94} Kenyon, S.~J., Dobrzycka, D., \& Hartmann, L.\ 1994, \aj, 108, 1872 
\bibitem[Kenyon \& Hartmann(1995)]{kh95} Kenyon, S.~J., \& Hartmann, L.\ 1995, \apjs, 101, 117 
\bibitem[Kenyon et al.(1998)]{kenyon+98} Kenyon, S.~J., Brown, D.~I., Tout, C.~A., \& Berlind, P.\ 1998, \aj, 115, 2491 
\bibitem[Kraus et al.(2011)]{kraus+11} Kraus, A.~L., Ireland, M.~J., Martinache, F., \& Hillenbrand, L.~A.\ 2011, \apj, 731, 8 
\bibitem[Luhman et al.(1997)]{llr97} Luhman, K.~L., Liebert, J., \& Rieke, G.~H.\ 1997, \apjl, 489, L165 
\bibitem[Luhman \& Rieke(1998)]{lr98} Luhman, K.~L., \& Rieke, G.~H.\ 1998, \apj, 497, 354 
\bibitem[Mathis(1990)]{mathis90} Mathis, J.~S.  1990, \araa, 28, 37 
\bibitem[Meeus et al.(2001)]{meeus+01} Meeus, G., Waters, L.~B.~F.~M., Bouwman, J., et al.\ 2001, \aap, 365, 476 
\bibitem[Mulders et al.(2011)]{mulders+11} Mulders, G.~D., Waters, L.~B.~F.~M., Dominik, C., et al.\ 2011, \aap, 531, A93 
\bibitem[Muzerolle et al.(1998)]{muzerolle98} Muzerolle, J., Hartmann, L., \& Calvet, N.\ 1998, \aj, 116, 2965 
\bibitem[Muzerolle et al.(2003)]{muzerolle03} Muzerolle, J., Calvet, N., Hartmann, L., \& D'Alessio, P.\ 2003, \apjl, 597, L149 
\bibitem[Monnier \& Millan-Gabet(2002)]{mmg02} Monnier, J.~D., \& Millan-Gabet, R.\ 2002, \apj, 579, 694 
\bibitem[Natta et al.(2001)]{natta+01} Natta, A., Prusti, T., Neri, R., Wooden, D., Grinin, V.~P., \& Mannings, V.\ 2001, \aap, 371, 186 
\bibitem[Najita et al.(2009)]{najita+09} Najita, J.~R., Doppmann, G.~W., Carr, J.~S., Graham, J.~R., \& Eisner, J.~A.\ 2009, \apj, 691, 738 
\bibitem[Pollack et al.(1994)]{pollack+94} Pollack, J.~B., Hollenbach, D., Beckwith, S., et al.\ 1994, \apj, 421, 615 
\bibitem[Posch et al.(2007)]{posch+07} Posch, T., Mutschke, H., Trieloff, M., \& Henning, T.\ 2007, \apj, 656, 615 
\bibitem[Rayner et al.(2003)]{ray03} Rayner, J. T., et al. 2003, \pasp, 115, 362
\bibitem[Rayner et al.(2009)]{rayner09} Rayner, J.~T., Cushing, M.~C., \& Vacca, W.~D.\ 2009, \apjs, 185, 289
\bibitem[Ricci et al.(2010)]{ricci+10} Ricci, L., Testi, L., Natta, A., et al.\ 2010, \aap, 512, A15 
\bibitem[Sargent et al.(2009)]{sargent+09} Sargent, B.~A., Forrest, W.~J., Tayrien, C., et al.\ 2009, \apjs, 182, 477 
\bibitem[Sembach \& Savage(1992)]{ss92} Sembach, K.~R., \& Savage, B.~D.\ 1992, \apjs, 83, 147 
\bibitem[Seperuelo Duarte et al.(2008)]{sd+08} Seperuelo Duarte, E., Alencar, S.~H.~P., Batalha, C., \& Lopes, D.\ 2008, \aap, 489, 349 
\bibitem[Shakura \& Sunyaev(1973)]{shakura_sunyaev73} Shakura, N.~I., \& Sunyaev, R.~A.\ 1973, \aap, 24, 337 
\bibitem[Siess et al.(2000)]{siess2000} Siess L., Dufour E., \& Forestini M.  2000, \aap, 358, 593
\bibitem[Tannirkulam et al.(2007)]{ajay+07} Tannirkulam, A., Harries, T.~J., \& Monnier, J.~D.\ 2007, \apj, 661, 374 
\bibitem[Tannirkulam et al.(2008)]{ajay+08} Tannirkulam, A., Monnier, J.~D., Millan-Gabet, R., et al.\ 2008, \apjl, 677, L51 
\bibitem[Vacca et al.(2003)]{vac03} Vacca, W. D., Cushing, M. C., \& Rayner J. T., 2003, \pasp, 115, 389
\bibitem[Vacca \& Sandell(2011)]{vacca11} Vacca, W.~D., \& Sandell, G.\ 2011, \apj, 732, 8 
\bibitem[Warren(1984)]{warren84} Warren, S.~G.\ 1984, \ao, 23, 1206 
\bibitem[Wendker(1995)]{wendker95} Wendker, H.~J.\ 1995, \aaps, 109, 177 
\bibitem[White \& Ghez(2001)]{wg01} White, R. J., \& Ghez, A. M.  2001, \apj, 556, 265
\bibitem[White \& Basri(2003)]{wb03} White, R.~J., \& Basri, G.\ 2003, \apj, 582, 1109 
\bibitem[White \& Hillenbrand(2004)]{wh04} White, R.~J., \& Hillenbrand, L.~A.\ 2004, \apj, 616, 998 
\bibitem[Zolensky et al.(2006)]{zolensky+06} Zolensky, M.~E., Zega, T.~J., Yano, H., et al.\ 2006, Science, 314, 1735 



\end{thebibliography}
\end{document}